\pdfoutput=1

\documentclass[11pt]{article}

\usepackage[preprint]{acl}

\usepackage{times}
\usepackage{latexsym}

\usepackage[T1]{fontenc}

\usepackage[utf8]{inputenc}

\usepackage{microtype}

\usepackage{inconsolata}

\usepackage{graphicx}

\usepackage{tabularray}
\usepackage{longtable}
\usepackage{xcolor}
\usepackage{colortbl}
\usepackage{multicol}
\usepackage{booktabs}
\usepackage{makecell}
\usepackage{amsmath, amssymb}
\usepackage{multirow}
\usepackage{mathtools}
\usepackage[inline]{enumitem}
\usepackage{subcaption}
\usepackage{float}
\usepackage{graphicx}
\usepackage{caption} 
\usepackage{upgreek}
\usepackage{seqsplit}
\usepackage{color,soul}
\usepackage{arydshln}
\usepackage{setspace}

\usepackage[most]{tcolorbox}
\newtcolorbox{mybox}[2][]{
    colback=white,
    colframe=gray!45,
    fonttitle=\bfseries,
    coltitle=black,
    sharp corners,
    title=#2,
    #1
}

\usepackage{amsmath,amsfonts,bm}

\def\eqref#1{equation~\ref{#1}}

\def\1{\bm{1}}

\def\vc{{\bm{c}}}
\def\vd{{\bm{d}}}

\def\vq{{\bm{q}}}

\def\vy{{\bm{y}}}

\DeclareMathAlphabet{\mathsfit}{\encodingdefault}{\sfdefault}{m}{sl}
\SetMathAlphabet{\mathsfit}{bold}{\encodingdefault}{\sfdefault}{bx}{n}

\usepackage{listings}
\usepackage{xcolor}

\usepackage{amssymb}
\usepackage{pifont}
\newcommand{\xmark}{\ding{55}}%
\newcommand{\ours}{CoLoR}

\definecolor{codegreen}{rgb}{0,0.6,0}
\definecolor{codegray}{rgb}{0.5,0.5,0.5}
\definecolor{codepurple}{rgb}{0.58,0,0.82}
\definecolor{backcolour}{rgb}{0.95,0.95,0.92}

\newcolumntype{a}{>{\columncolor{ggr}}c}

\definecolor{ggr}{gray}{0.92}
\definecolor{gg}{HTML}{E0FEFE}

\lstdefinestyle{mystyle}{
    backgroundcolor=\color{backcolour},   
    commentstyle=\color{codegreen},
    keywordstyle=\color{magenta},
    numberstyle=\tiny\color{codegray},
    stringstyle=\color{codepurple},
    basicstyle=\ttfamily\footnotesize,
    breakatwhitespace=false,         
    breaklines=true,                 
    captionpos=b,                    
    keepspaces=true,                 
    numbers=left,                    
    numbersep=5pt,                  
    showspaces=false,                
    showstringspaces=false,
    showtabs=false,                  
    tabsize=2
}

\usepackage{caption}
\usepackage{subcaption}
\title{Efficient Long Context Language Model Retrieval with Compression}

\author{
    Minju Seo$^{1}$ \;\;
    Jinheon Baek$^{1}$ \;\;
    Seongyun Lee$^{1}$ \;\;
    Sung Ju Hwang$^{1,2}$ \\
    KAIST$^{1}$, \;\; DeepAuto$^{2}$ \\
    \texttt{\{minjuseo, jinheon.baek, seongyun, sungju.hwang\}@kaist.ac.kr}
}

\begin{document}

\maketitle
\begin{abstract}
Long Context Language Models (LCLMs) have emerged as a new paradigm to perform Information Retrieval (IR), which enables the direct ingestion and retrieval of information by processing an entire corpus in their single context, showcasing the potential to surpass traditional sparse and dense retrieval methods. However, processing a large number of passages within in-context for retrieval is computationally expensive, and handling their representations during inference further exacerbates the processing time. To address this challenge, we aim to make LCLM retrieval more efficient and potentially more effective with passage compression. Specifically, we propose a new compression approach tailored for LCLM retrieval, which is trained to maximize the retrieval performance while minimizing the length of the compressed passages. To accomplish this, we generate the synthetic data, where compressed passages are automatically created and labeled as chosen or rejected according to their retrieval success for a given query, and we then train the proposed \textbf{Co}mpression model for \textbf{Lo}ng context \textbf{R}etrieval (CoLoR) with this data via preference optimization while adding the length regularization loss on top of it to enforce brevity. We perform extensive experiments on nine datasets, and showcase that CoLoR improves the retrieval performance by 6\% while compressing the in-context size by a factor of 1.91. Our code is available at: \url{https://github.com/going-doer/CoLoR}.
\end{abstract}

\section{Introduction}

The context size of Language Models (LMs) refers to the maximum number of tokens that the model can process in a single input, which has rapidly expanded, growing from a few hundred to 128K, and recently reaching 1M tokens in Long Context Language Models (LCLMs)~\cite{GPT4, Gemini, Claude3}. Notably, this expansion has unlocked new capabilities, enabling models to handle tasks that require extensive context lengths, such as summarization or question answering over long articles~\cite{Xu2023RetrievalML, FABLES}. In addition to this, LCLMs go beyond these relatively simple tasks to handle more complex tasks, such as Information Retrieval (IR) or Text-to-SQL, which not only demand long-range context understanding but also involve reasoning across multiple documents or structured queries~\cite{Lee2024LOFT, LEval, Liu2023LostIT, Xu2023RetrievalML}. Furthermore, due to their impressive performance, they have established a new paradigm in LM utilization and task solving. For example, in IR tasks that we focus on, LCLMs are capable of processing an entire corpus with a large number of documents along with a user query in their single context, leading to more precise identification of relevant information and further surpassing traditional sparse or dense retrieval approaches in many cases~\cite{Lee2024LOFT}. Yet, LCLMs face the limitation that the required computational resources scale with the input length, which has been overlooked by existing work. 

\begin{figure*}[t!]
    \centering
    \includegraphics[width=0.975\linewidth]{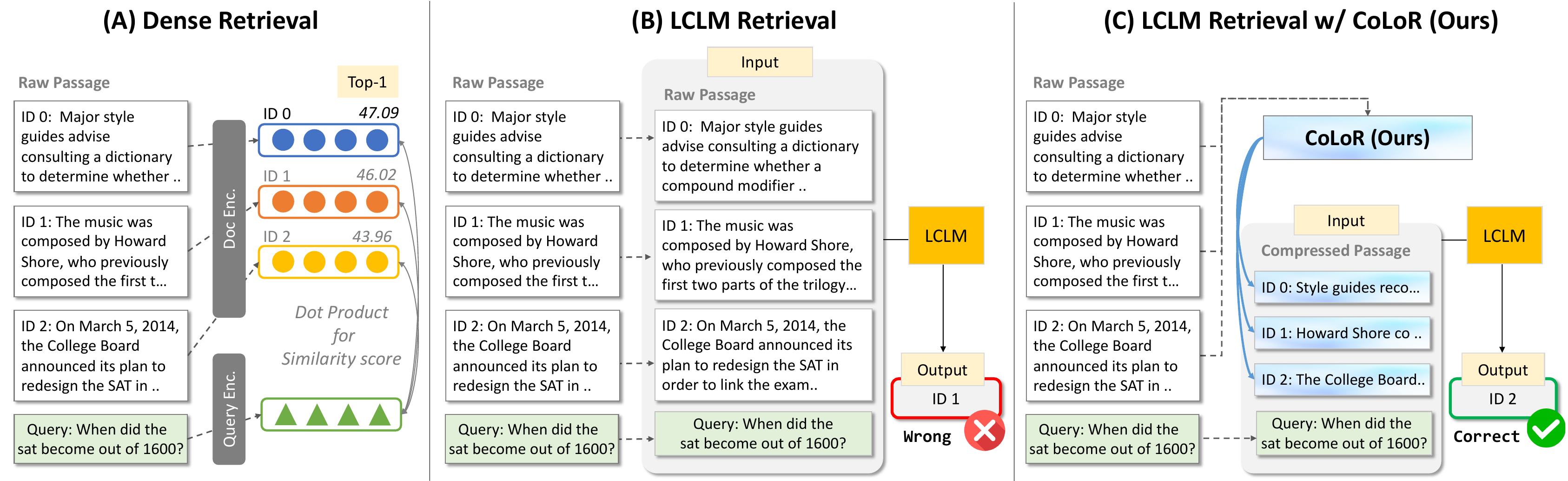}
    \vspace{-0.05in}
    \caption{Comparison of different IR approaches. \textbf{(A) Dense Retrieval.} To identify relevant documents to the given query, it first embeds them into the vector space and then calculates their semantic similarity. \textbf{(B) LCLM Retrieval.} The LCLM takes and processes the raw passages from the corpus along with the query in the input context, and identifies the relevant passages. \textbf{(C) CoLoR.} We compress the raw passages, and use the compressed passages alongside the query as the LCLM input for retrieval.}
    \label{fig:concept}
    \vspace{-0.1in}
\end{figure*}

To tackle this challenge, we propose a more efficient (and potentially more effective) method for LCLM retrieval. Specifically, instead of relying on the original passages, our approach uses compressed passages that retain the core information while filtering our irrelevant details, leading to a substantial reduction in input size. It is worth noting that, while there are existing compression methods~\cite{Jiang2023LLMLinguaCP, Xu2024RECOMPIR} particularly designed for text generation or retrieval-augmented generation, they are largely suboptimal for retrieval since they are not trained to prioritize key elements crucial for precise retrieval, such as relevance to the query or fine-grained document distinctions. In contrast, our compression model is trained to optimize the LCLM retrieval performance, while minimizing the passage length with the regularization term added on top of the optimization objective.

Specifically, to train our compression model, we leverage the strategy of Odds Ratio Preference Optimization (ORPO)~\cite{ORPO}, a method well-suited for maximizing the preference by comparing pairs of samples and learning to prefer one over the other based on the specific objective. We note that this is particularly useful for our scenario since it allows us to rank compressed passages according to their retrieval performance, helping the model distinguish between more and less effective compressions, without the need for generating the ground-truth compression outputs manually. In other words, to generate the training data for it, we automatically create multiple compressed versions of each raw passage in the corpus and evaluate their retrieval success with the LCLM retrieval outcome. Subsequently, compressed samples are labeled as either chosen or rejected based on this evaluation, which can ultimately guide the model toward generating the effective compression for passages for retrieval during preference optimization. However, using ORPO alone might not sufficiently reduce the output length of the compressed passages. Thus, to overcome this, we further introduce a dynamic regularization term that adjusts the odds ratio loss in the training objective based on the length difference between chosen and rejected samples (where we additionally consider the compressed passages with the correct retrieval as rejected if there exist more shorter ones with the correct retrieval), which can encourage the model to prioritize brevity while at the same time optimizing the retrieval accuracy. We refer to our approach as \textbf{Co}mpression for \textbf{Lo}ng Context Language Model \textbf{R}etrieval (\textbf{\ours}), illustrated in Figure~\ref{fig:concept} with previous IR methods.

We experimentally validate the effectiveness of CoLoR on 9 datasets for LCLM retrieval, including 5 single-document and 4 multi-document retrieval scenarios. On a battery of tests conducted, we then demonstrate that our approach not only improves the retrieval performance by 6\% but also reduces the context length for retrieval by a factor of 1.91 over the baseline that uses the original passages, and it is further superior to the existing text compression (or summarization) methods. Further, we show that our compression method can be generalizable to datasets not seen during its training. 

\section{Related Work}

\paragraph{Information Retrieval} The goal of Information Retrieval (IR) is to fetch documents relevant to a query, which has evolved significantly with their application to various tasks, such as web search and question answering. Early approaches used sparse retrieval methods, such as BM25~\cite{BM25}, which are based on lexical matching between queries and documents. As the field progressed, dense retrieval techniques have developed, leveraging text embedding models to capture richer semantic relationships between queries and documents. Notable examples of dense retrievers include SentenceBERT~\cite{SentenceBERT}, Dense Passage Retrieval (DPR)~\cite{DPR}, and SentenceT5~\cite{SentenceT5}. More recently, researchers have begun transforming Large Language Models (LLMs) into retrieval systems~\cite{LLM2Vec}, which aim to utilize the vast contextual understanding capability of LLMs in representing documents and queries. Following this line of approaches, our work extends by utilizing LCLMs as the retrieval mechanism.

\paragraph{Long Context Language Models}
The recent expansion in context length of LLMs, which is called Long Context Language Models (LCLMs), has empowered them to process and comprehend much larger amounts of information. Specifically, models like YaRN~\cite{YaRN}, Longformer~\cite{Longformer}, Gemini~\cite{Gemini}, GPT-4~\cite{GPT4}, and Claude~\cite{Claude3} exemplify this advancement. In tandem with these developments, new benchmarks have been introduced to assess the capabilities of LCLMs across various tasks. For example, many studies~\cite{Bai2023LongBenchAB, Li2023LooGLECL, Liu2023LostIT, Yuan2024LVEvalAB, Wang2024AdaLEvalEL} evaluated their performance in long-context understanding, and there are also other studies that focused on more specialized areas such as code comprehension~\cite{Liu2024RepoQAEL} and training-free in-context learning~\cite{Bertsch2024InContextLW, Li2024LongcontextLS}. Moreover, \citet{Lee2024LOFT} demonstrated that LCLMs outperform traditional fine-tuned specialized models in several areas (such as IR). However, despite these promising results, \citet{Liu2023LostIT} highlighted the persistent challenges LCLMs face in fully grasping complex long contexts with high computational costs. In contrast to existing work that has mainly explored the potential and diverse applications of LCLMs, we take a different direction on improving the efficiency of LCLMs in the context of IR by reducing the context size while maintaining or enhancing performance.

\paragraph{Prompt Compression}
As LCLMs handle increasingly longer contexts, the corresponding rise in computational costs has sparked research into methods for prompt compression. Extractive compression is one common approach, where only the relevant tokens are retained. This often involves techniques such as token pruning, which require assessing the importance of individual tokens based on specific metrics, for example, utilizing the self-information or perplexity of the model~\cite{Jiang2023LLMLinguaCP, Li2023CompressingCT}. However, these methods typically require access to the model's internal processes, making them feasible only for white-box models. In contrast, abstractive compression methods generate the condensed prompts without needing to preserve the original token order or structure, and can be applied to both black-box and white-box models as they do not rely on internal model access. For example, \citet{Xu2024RECOMPIR} generate compressed content from multiple documents in Retrieval Augmented Generation (RAG) settings, and \citet{Wang2023LearningTF} use a similar approach to generate distilled documents. Despite these advancements, previous work focusing on context compression in RAG or instruction-following tasks is not well-optimized for retrieval tasks, as it does not cater specifically to the needs of retrieval. To address this gap, we propose an abstractive compression model designed to improve the efficacy of LCLM retrieval.

\paragraph{Preference Optimization}
Aligning the language models with human preferences has become a key focus in improving response generation~\cite{RLHF, Zhao2023SLiCHFSL, DPO, ORPO}. A prominent approach is Reinforcement Learning from Human Feedback (RLHF)~\cite{RLHF, RLHF2}, which leverages a reinforcement learning framework where a policy model learns to evaluate and choose actions based on the state of an environment, with human feedback acting as a reward signal. Notably, what sets these approaches apart is their ability to train models using only preference selections on outputs, without needing explicit ground truth answers. Additionally, Odds Ratio Preference Optimization (ORPO)~\cite{ORPO} further simplifies this by removing the requirement for a reference model during training, allowing single-step learning via preference selection of outputs. In our approach, we apply this framework to train the compression model without needing ground truth labels, enabling the model to learn based on a pair of compression outputs and their retrieval results.

\section{Methodology}

\begin{figure*}[t!]
    \centering
    \includegraphics[width=0.975\linewidth]{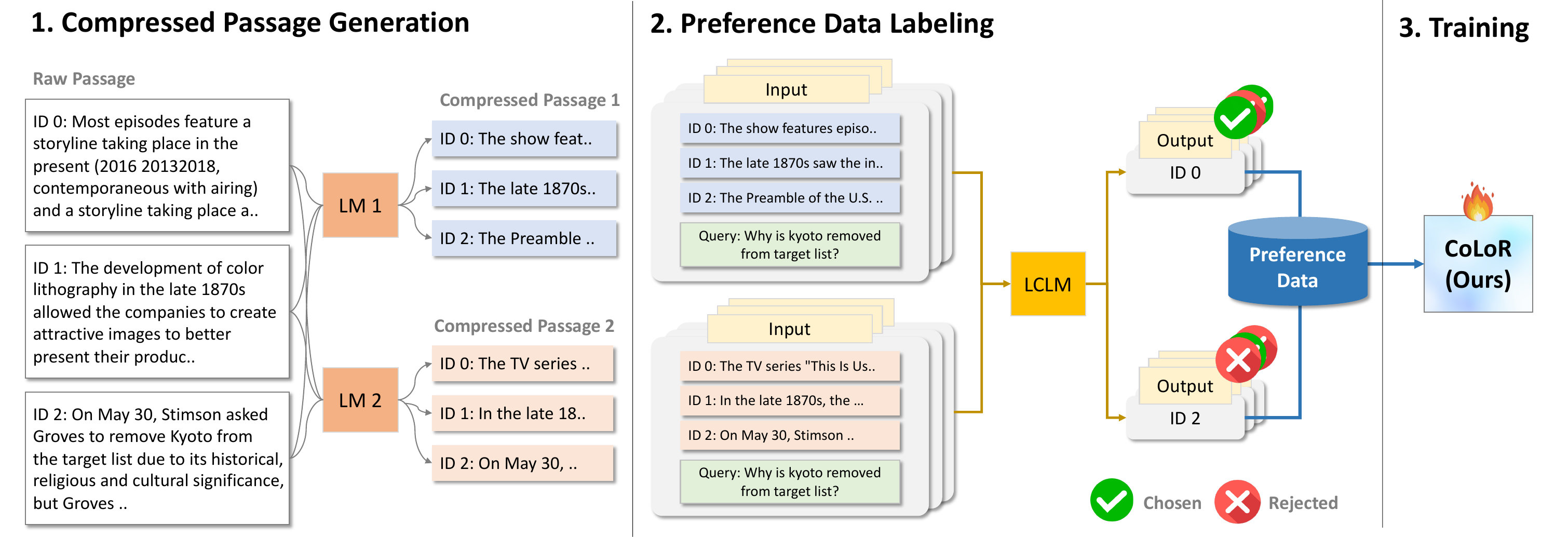}
    \vspace{-0.05in}
    \caption{\textbf{Overview of Training Processes for CoLoR}. 1. We first create the training data for CoLoR by generating multiple compressed passages from their original passages with multiple LMs. 2. The compressed passages and their associated query are used as input to the LCLM, and their retrieval performance is measured to label them as either chosen or rejected based on retrieval results. 3. CoLoR is trained using the pairs of chosen and rejected compressed passages obtained from previous steps.}
    \label{fig:overview}
\end{figure*}

\subsection{Problem Statement} 
We begin with formally explaining IR and LCLMs.

\paragraph{Information Retrieval}
In a typical IR task, given a query $\vq$, its objective is to retrieve a ranked list $k$ relevant entries from a corpus $\mathcal{C}$, formulated as follows: $\left\{ \vd_i \right\}_{i=1}^{k} = \texttt{Retriever}(\vq, \mathcal{C})$, where $\vd_i$ is a document from $\mathcal{C}$. The query $\vq$ is typically textual, and $\mathcal{C}$ is a collection of documents. Traditionally, $\texttt{Retriever}$ is operationalized with the sparse retrieval based on lexical term matching~\cite{BM25} or the neural embedding-based dense methods~\cite{DPR}.

\paragraph{Long Context Language Model Retrieval}
Recently, Long Context Language Models (LCLMs) have emerged with the ability to process extended contexts, enabling them to handle inputs spanning dozens of documents. This capability has given rise to a new paradigm called LCLM Retrieval, which utilizes LCLMs to solve IR tasks~\cite{Lee2024LOFT}. To be formal, similar to the typical IR approaches, LCLM retrieval aims to retrieve relevant documents from the corpus $\mathcal{C}$ for the query $\vq$, which can be represented as follows: $\left\{ \vd_i \right\}_{i=1}^{k} = \texttt{LCLM}(\mathcal{T}(\vq, \mathcal{C}))$, where $\mathcal{T}$ is the prompt template which serves as the structured format that outlines the context for LCLMs (including task descriptions) to direct them in performing retrieval. It is worth noting that, unlike traditional retrieval methods (sparse or dense), which involve pairing each document with a query and calculating similarity scores to rank documents, LCLM retrieval takes both the entire corpus and the query as the single input and directly identifies relevant documents within it. However, a significant challenge with LCLM retrieval lies in the use of raw documents (or passages) as input, as they often contain unnecessary, redundant, and irrelevant context, leading to increased computational costs.

\subsection{CoLoR: Compression for Long~Context~Language~Model~Retrieval}
To tackle the inefficiency of using raw passages in LCLM retrieval, we propose using compressed passages to reduce the computational overhead without compromising retrieval effectiveness. There are two common approaches to compressing passages: using prompt-based methods with LLMs or leveraging off-the-shelf compression models. However, prompt-based methods often fail to achieve optimal compression ratios and are not tailored to enhance IR performance. Likewise, existing compression models, which are typically designed for tasks other than IR (such as RAG or instruction following), are not well-suited for LCLM retrieval.

To address these limitations, we propose CoLoR (Compression for Long Context Retrieval), which is a novel compression model designed for LCLM retrieval. CoLoR generates compressed passages by learning to balance two objectives: maintaining high retrieval accuracy and reducing passage length. To achieve this, we leverage preference optimization using synthetic preference data, where compressed passages are automatically generated, and labeled based on their retrieval success as well as their resulting lengths. This allows the model to distinguish between more and less effective compressions without manually collecting labels.

Formally, let $\vd_i \in \mathcal{C}$ represent a raw document (or passage) from the corpus. Our goal is to apply the CoLoR model to compress each document $\vd_i$ into a more concise representation $\vc_i = \texttt{CoLoR}(\vd_i)$, where $\vc_i$ is the compressed version of the raw document. Ideally, the compressed passage $\vc_i$ retains the most relevant information while filtering out unnecessary details, therefore, reducing the length of the input to the LCLM. After compressing every document in the corpus, during the retrieval process, instead of directly using the original corpus $\mathcal{C}$, the LCLM ingests the compressed corpus $\mathcal{C}^* = \{ \vc_i \}_{i=1}^{|\mathcal{C}|}$, where each element in $\mathcal{C}$ is transformed by CoLoR. In other words, the retrieval process can be redefined as follows: $\left\{ \vd_i \right\}_{i=1}^{k} = \texttt{LCLM}(\mathcal{T}(\vq, \mathcal{C}^*))$ with $|\mathcal{T}(\vq, \mathcal{C}^*)| \ll |\mathcal{T}(\vq, \mathcal{C})|$ where $|\cdot|$ measures the number of tokens in the resulting prompt.

\subsubsection{Training Recipe for CoLoR}
We now turn to explaining the details of how we train our CoLoR to optimize efficiency while improving retrieval accuracy, illustrated in Figure~\ref{fig:overview}.

\paragraph{Data Collection}
To train CoLoR for LCLM retrieval, we need to create a new dataset as no such datasets are available. Our data creation process begins with leveraging multiple LLMs to generate multiple compressed versions of raw documents (or passages), by prompting them with the prompt template: $\mathcal{T} =$ \texttt{Summarize the following content: \{passage\}}, formalized as follows: $\vc = \texttt{LLM}(\mathcal{T}(\vd))$ where $\vc$ is the compressed passage and $\vd$ is its original version. After that, the compressed passages are used as inputs for the LCLM retrieval, which are then labeled as either chosen or rejected based on two criteria: 1) whether the compressed passage is correctly retrieved in response to its associated query and 2) whether its length is shorter than any of the other successfully retrieved compressed passages. For instance, if several compressed versions of a passage are retrieved correctly, the shortest of these is labeled as chosen, while the others are labeled as rejected. Also, if the retrieval with the compressed passage fails, it is labeled as rejected. This allows us to create the dataset with pairs of chosen and reject compression results for training CoLoR with preference optimization, which can ultimately prioritize compressions that improve retrieval accuracy while minimizing passage length.

\paragraph{CoLoR Optimization}
To optimize our compression model, we leverage the Odds Ratio Preference Optimization (ORPO)~\cite{ORPO}, an approach designed for training models by comparing pairs of chosen and rejected samples without the need for a reference model. ORPO is particularly suited for our task, as it allows us to directly optimize the model to prefer compressed passages that yield better retrieval performance with the shortest length. Formally, the standard ORPO loss function measures the odds ratio between the likelihood of generating the chosen response $\vy_w$ and the rejected response $\vy_l$, represented as follows:
\begin{equation}
\label{eqa:orpo}
\mathcal{L}_{\text{ORPO}} = \mathbb{E}_{(\vq,\vy_w,\vy_l)} \left[ \mathcal{L}_{\text{SFT}} + \lambda \cdot \mathcal{L}_{\text{OR}} \right], \nonumber
\end{equation}
where $\mathcal{L}_{\text{SFT}}$ is the supervised fine-tuning loss for the chosen response based on the causal language modeling negative log-likelihood, and $\mathcal{L}_{\text{OR}}$ is the loss for the odds ratio of the chosen response over the rejected one (See~\citet{ORPO} for details). 

However, while ORPO enables effective preference learning for making the compression model, it does not inherently reduce the length of the compressed passages as much as desired. To overcome this, we further propose to use a dynamic regularization term that adjusts the odds ratio loss based on the length difference between rejected and chosen samples. Specifically, we redefine the ORPO loss by multiplying it with a specific factor determined as the length difference between $\vy_l$ and $\vy_w$ (where $\vy_l$ is always longer than $\vy_w$ based on the criteria in our data collection process), as follows:
\vspace{-0.075in}

{
\fontsize{10.0pt}{10.0pt}\selectfont
\begin{align}
\mathcal{L}_{\text{CoLoR}} = \mathbb{E}_{(\vq,\vy_w,\vy_l)} \left[ \mathcal{L}_{\text{SFT}} + \lambda \cdot \mathcal{L}_{\text{OR}} \cdot \left( |\vy_l| - |\vy_w| \right) \right], \nonumber
\end{align}
\fontsize{10.0pt}{10.0pt}\selectfont
}

\noindent
where $|\cdot|$ measures the length of the compressed passage. This extra regularization term directly allows the model to make larger updates in cases where the chosen sample is significantly shorter than the rejected one, making it to favor concise outputs without sacrificing retrieval accuracy.

\begin{table*}[t]
\caption{\textbf{Results on LCLM retrieval}. \textbf{Type} refers to a compression type: \xmark~ denotes no compression, Ex. denotes extractive compression, and Ab. denotes abstractive compression. \dag~denotes multi-document retrieval. Comp. is the compression rate.}
\vspace{-0.05in}
\label{tab:main_result}
\small
\centering
\resizebox{0.975\textwidth}{!}{
\begin{tabular}{lccccccccccc}
\toprule
& & \multicolumn{2}{c}{FEVER} & \multicolumn{2}{c}{FIQA} & \multicolumn{2}{c}{MS MARCO} & \multicolumn{2}{c}{NQ} & \multicolumn{2}{c}{SciFact} \\

\cmidrule(l{2pt}r{2pt}){3-4} \cmidrule(l{2pt}r{2pt}){5-6} \cmidrule(l{2pt}r{2pt}){7-8} \cmidrule(l{2pt}r{2pt}){9-10} \cmidrule(l{2pt}r{2pt}){11-12}  
 \textbf{Methods} & \textbf{Type} & R@1 & Comp. & R@1 & Comp. & R@1 & Comp. & R@1 & Comp.& R@1 & Comp. \\

 \midrule
 \midrule
 
 Raw Passage & \xmark & 0.95 & 1.00x & 0.63 & 1.00x & 0.90 & 1.00x & 0.97 & 1.00x & 0.57 & 1.00x \\
 Document Title& \xmark & 0.97 & 31.65x & N/A & N/A & N/A & N/A & 0.86 & 22.56x & 0.68 & 13.98x \\

\noalign{\vskip 0.25ex}\cdashline{1-12}\noalign{\vskip 0.75ex}
BM25& \xmark & 0.93 & 1.00x & 0.41 & 1.00x & 0.78 & 1.00x & 0.79 & 1.00x & 0.75 & 1.00x \\
DPR & \xmark  & 0.89 & 1.00x & 0.31 & 1.00x & 0.85 & 1.00x & 0.94 & 1.00x & 0.35 & 1.00x \\
 
  \midrule
  
 Selective Context (0.6)& Ex. & 0.96 & 1.83x & 0.35 & 1.83x & 0.52 & 1.92x & 0.95 & 1.88x & 0.71 & 1.89x \\
 Selective Context (0.3)& Ex.  & 0.95 & 4.58x & 0.15 & 4.24x & 0.19 & 4.69x & 0.90 & 4.91x & 0.63 & 5.33x \\
 LLMLingua & Ex. & 0.96 & 1.70x & 0.46 & 1.6x & 0.83 & 1.12x & 0.97 & 1.24x & 0.74 & 1.92x \\
  
 \noalign{\vskip 0.25ex}\cdashline{1-12}\noalign{\vskip 0.75ex}
  
 Comp. w/ GPT & Ab.  & 0.96 & 1.74x & 0.68 & 2.09x & 0.92 &  1.39x & 0.99 & 1.37x & 0.73 & 1.71x \\
 Comp. w/ Phi & Ab. & 0.95 & 1.83x & 0.68 & 2.31x & 0.92 & 1.41x & 0.99 & 1.39x & 0.71 & 1.78x \\
 RECOMP & Ab. & 0.96 & 3.02x & 0.36 & 2.00x & 0.75 & 1.77x & 0.98 & 1.97x & 0.70 & 2.31x \\
 COMPACT & Ab. & 0.96 & 2.21x & 0.48 & 2.56x & 0.88 & 1.17x & 0.98 & 1.32x & 0.77 & 2.95x   \\

 \noalign{\vskip 0.25ex}\cdashline{1-12}\noalign{\vskip 0.75ex}

 \cellcolor{gg}CoLoR (Ours) & \cellcolor{gg} Ab.& \cellcolor{gg} 0.94 & \cellcolor{gg} 2.15x & \cellcolor{gg} 0.73 & \cellcolor{gg} 2.82x & \cellcolor{gg} 0.95 & \cellcolor{gg} 1.63x & \cellcolor{gg} 0.98 & \cellcolor{gg} 1.62x & \cellcolor{gg} 0.75 & \cellcolor{gg} 2.12x \\
\midrule

\midrule

 & & \multicolumn{2}{c}{HotPotQA\textsuperscript{\dag}} & \multicolumn{2}{c}{MuSiQue\textsuperscript{\dag}} & \multicolumn{2}{c}{QAMPARI\textsuperscript{\dag}} & \multicolumn{2}{c}{QUEST\textsuperscript{\dag}} & \multicolumn{2}{c}{Average} \\
\cmidrule(l{2pt}r{2pt}){3-4} \cmidrule(l{2pt}r{2pt}){5-6} \cmidrule(l{2pt}r{2pt}){7-8} \cmidrule(l{2pt}r{2pt}){9-10} \cmidrule(l{2pt}r{2pt}){11-12}
 \textbf{Methods} & \textbf{Type} & F1@2 & Comp. & F1@5 & Comp. & F1@5 & Comp.& F1@3 & Comp.& Perf. & Comp. \\
\midrule
 \midrule
 
 Raw Passage & \xmark & 0.87 & 1.00x & 0.35 & 1.00x & 0.56 & 1.00x & 0.33 & 1.00x & \cellcolor{ggr} 0.68 & \cellcolor{ggr} 1.00x \\
 Document Title & \xmark & 0.65 & 12.52x & 0.38 & 23.43x & 0.25 & 22.48x & 0.11 & 55.65x & \cellcolor{ggr} 0.60 &  \cellcolor{ggr} 26.04x \\
 \noalign{\vskip 0.25ex}\cdashline{1-12}\noalign{\vskip 0.75ex}
 BM25 & \xmark & 0.82 & 1.00x & 0.39 & 1.00x & 0.76 & 1.00x & 0.37 & 1.00x & \cellcolor{ggr} 0.64 & \cellcolor{ggr} 1.00x \\
 DPR & \xmark & 0.79 & 1.00x & 0.46 & 1.00x & 0.55 & 1.00x & 0.31 & 1.00x & \cellcolor{ggr} 0.61 & \cellcolor{ggr} 1.00x  \\
 
  \midrule

 Selective Context (0.6) & Ex. & 0.79 & 2.06x & 0.37 & 1.93x & 0.42 & 1.95x & 0.19 & 2.01x & \cellcolor{ggr} 0.58 & \cellcolor{ggr} 1.92x \\
 Selective Context (0.3) & Ex. & 0.68 & 5.49x & 0.37 & 5.10x & 0.30 & 5.30x & 0.11 & 5.41x & \cellcolor{ggr} 0.48 & \cellcolor{ggr} 5.01x \\
 LLMLingua & Ex.  & 0.81 & 1.13x & 0.36 & 1.23x & 0.54 & 1.05x & 0.21 & 1.75x & \cellcolor{ggr} 0.65 & \cellcolor{ggr} 1.42x \\
 
 \noalign{\vskip 0.25ex}\cdashline{1-12}\noalign{\vskip 0.75ex}
 Comp. w/ GPT & Ab. & 0.87 & 1.20x & 0.40 & 1.32x & 0.54 & 1.28x & 0.32 & 1.95x & \cellcolor{ggr} 0.71 & \cellcolor{ggr} 1.56x \\
 Comp. w/ Phi & Ab. & 0.85 & 1.21x & 0.41 & 1.39x & 0.55 & 1.33x & 0.31 & 2.03x & \cellcolor{ggr} 0.71 & \cellcolor{ggr} 1.63x \\

 RECOMP & Ab. & N/A & 0.76x & 0.38 & 1.00x & 0.53 & 0.97x & 0.21 & 1.50x & \cellcolor{ggr} 0.54 & \cellcolor{ggr} 1.70x \\

 COMPACT & Ab. & 0.87 & 1.14x & 0.37 & 1.47x & 0.52 & 1.51x & 0.30 & 3.47x & \cellcolor{ggr} 0.68 & \cellcolor{ggr} 1.98x  \\

 \noalign{\vskip 0.25ex}\cdashline{1-12}\noalign{\vskip 0.75ex}

 \cellcolor{gg} CoLoR (Ours) & \cellcolor{gg} Ab. & \cellcolor{gg} 0.86 & \cellcolor{gg} 1.37x & \cellcolor{gg} 0.42 & \cellcolor{gg} 1.55x & \cellcolor{gg} 0.55 & \cellcolor{gg} 1.50x & \cellcolor{gg} 0.33 & \cellcolor{gg} 2.39x & \cellcolor{gg} 0.72 & \cellcolor{gg} 1.91x \\
 
\bottomrule

\end{tabular}
}
\vspace{-0.05in}
\end{table*}

\section{Experiment Setup}

\subsection{Datasets}
We evaluate the performance of CoLoR on 9 widely used LCLM retrieval benchmark datasets, following the setup from~\citet{Lee2024LOFT}, including 5 single-document retrieval datasets: FEVER, FIQA, MS MARCO, NQ, and SciFact~\cite{FEVER, FIQA, MSMARCO, NQ, SciFact} and 4 multi-document retrieval datasets: HotpotQA, MuSiQue, QAMPARI, and QUEST~\cite{HotpotQA, MuSiQue, QAMPARI, QUEST}. Note that single-document retrieval tasks involve retrieving a single document relevant to a query, whereas multi-document retrieval tasks require retrieving two or more documents. We provide the detailed statistics in Table~\ref{tab:dataset_stat}.

\subsection{Baselines and Our Model}
We evaluate CoLoR against baselines, as follows:
\begin{enumerate*}[itemsep=0.0mm, parsep=1pt, leftmargin=*]
    \item \textbf{Raw Passage} is a standard approach for LCLM retrieval that directly uses raw passages;
    \item \textbf{Document Title} uses only the titles of the passages without the full content;
    \item \textbf{Zero-Shot Compression} uses LLMs to compress passages via prompting, for example, $\mathcal{T} =$ \texttt{Summarize the following content: \{passage\}}, with GPT-4o-mini~\cite{GPT4} and Phi3~\cite{phi3};
    \item \textbf{Selective Context} is an extractive compression method that selects tokens based on the self-information of the model~\cite{Li2023CompressingCT}, where we use two compression rates: 0.3 and 0.6;
    \item \textbf{LLMLingua} is an extractive compression method that selects tokens based on the perplexity scores~\cite{Jiang2023LLMLinguaCP};
    \item \textbf{RECOMP} is an abstractive compression method that compresses multiple documents designed for RAG scenarios~\cite{Xu2024RECOMPIR};
    \item \textbf{COMPACT} is an abstractive compression method that compresses and refines passages iteratively for question answering~\cite{Yoon2024CompActCR};
    \item \textbf{CoLoR (Ours)} is our abstractive compression method, trained to maximize the retrieval accuracy and minimize the compressed passage length, with the preference optimization.
\end{enumerate*}
Additionally, we also include BM25~\cite{BM25} (a sparse retriever that scores documents based on term frequency and inverse document frequency) and DPR~\cite{DPR} (a dense retriever that uses embeddings to match queries and relevant passages) as the reference to the performance of conventional retrievers, which are neither comparable nor our competitors.

\subsection{Evaluation Metrics}
For single-document retrieval, we evaluate performance with \textbf{Recall@1 (R@1)}, which measures the proportion of queries for which the top-ranked document is relevant. For multi-document, we use \textbf{F1@$k$}, a metric combining \textbf{Precision@$k$} (the proportion of correctly retrieved relevant documents in the top $k$ results) and \textbf{Recall@$k$} (the proportion of relevant documents retrieved from up to $k$ total). For compression efficiency, we compute the \textbf{compression rate (Comp.)}, defined as the average number of tokens in raw passages divided by the average number of tokens in compressed passages. 

\subsection{Implementation Details}
To ensure a fair comparison across all experiments, we use GPT-4o-mini as the underlying LCLM. We use the Phi-3-mini-4k-instruct model as the base model for our compression method, CoLoR. For the prompt, we structure it as a sequence of the corpus, 5-shot examples, and query, following~\citet{Lee2024LOFT}. Additional details are in Appendix~\ref{appendix:setups}.

\begin{table}[t]
\caption{\textbf{Results on out-of-domain datasets}, where the target retrieval category is excluded from the training of CoLoR*.}
\vspace{-0.05in}
\label{tab:ood_results}
\small
\centering
\resizebox{0.475\textwidth}{!}{
\begin{tabular}{lccccccc}
\toprule
  & \multicolumn{2}{c}{Raw Passage} &  \multicolumn{2}{c}{Comp w/ Phi} &  \multicolumn{2}{c}{CoLoR*} \\

 \cmidrule(l{2pt}r{2pt}){2-3} \cmidrule(l{2pt}r{2pt}){4-5}  \cmidrule(l{2pt}r{2pt}){6-7}  
 & Perf. & Comp. & Perf. & Comp. & Perf. & Comp.  \\
 \midrule
 \midrule
 \multicolumn{4}{l}{\textit{Fact-checking}} \\ 
 FEVER & 0.95 & 1.00x & 0.95 & 1.83x & 0.95 & 2.14x  \\
 SciFact & 0.71 & 1.00x & 0.71 & 1.78x & 0.73 & 2.13x  \\
 \noalign{\vskip 0.25ex}\cdashline{1-7}\noalign{\vskip 0.75ex} 
 Average & 0.83 & 1.00x & 0.83 & 1.81x & 0.84 & 2.14x \\
\midrule
 \multicolumn{4}{l}{\textit{Multi-document}} \\ 
 HotpotQA & 0.85 & 1.00x & 0.85 & 1.21x & 0.87 & 1.36x\\
 MuSiQue & 0.35 & 1.00x & 0.41 & 1.39x & 0.40 & 1.54x\\
 QAMPARI & 0.56 & 1.00x & 0.55 & 1.33x & 0.56 & 1.48x\\
 QUEST & 0.33 & 1.00x & 0.31 & 2.03x & 0.32 & 2.33x\\
\noalign{\vskip 0.25ex}\cdashline{1-7}\noalign{\vskip 0.75ex}
Average & 0.52 & 1.00x & 0.53 & 1.49x & 0.54 & 1.68x \\
\midrule
 \multicolumn{4}{l}{\textit{Argument}} \\ 
 ArguAna & 0.28 & 1.00x & 0.27 & 2.26x & 0.34 & 2.73x \\
 Touché-2020 & 0.76 & 1.00x & 0.79 & 3.79x & 0.79 & 4.66x \\
 \noalign{\vskip 0.25ex}\cdashline{1-7}\noalign{\vskip 0.75ex}
Average & 0.52 & 1.00x & 0.53& 3.03x & 0.57 & 3.70x \\
\bottomrule

\end{tabular}
}
\vspace{-0.05in}
\end{table}

\section{Experiment Results}

\paragraph{Main Results}
We report main results in Table~\ref{tab:main_result}, demonstrating that the proposed CoLoR approach consistently outperforms all baseline methods on both the single-document and multi-document retrieval tasks while at the same time substantially compressing the input context size of LCLMs for retrieval. Specifically, CoLoR achieves a compression rate that reduces the input size by a factor of 1.91, while also improving retrieval performance by 6\%, compared to the standard approach with raw passages. Also, our CoLoR provides the superior quality compressed passages for retrieval, compared to extractive and abstractive compression models. For instance, when compared with extractive methods (Selective Context and LLMLingua), CoLoR consistently demonstrates better retrieval performance. In addition to this, even against the strong proprietary and open-source models (such as GPT and Phi3), CoLoR excels, particularly in terms of the compression rate, highlighting the limitations of relying on prompting techniques to generate compressed passages for retrieval. Lastly, when compared with multi-document compression methods such as RECOMP and COMPACT, our CoLoR significantly outperforms them in the retrieval performance with the similar compression rate, which further confirms the necessity of task-specific training for passage compression for LCLM retrieval.

\paragraph{Results on Out-of-Domain Datasets}
To assess the generalizability of our compression approach (CoLoR) on datasets not seen for training, we evaluate its performance in out-of-domain settings by excluding a set of datasets from each retrieval category (such as fact-checking, multi-document, and argument) from the training process and testing on them. As shown in Table~\ref{tab:ood_results}, we observe that CoLoR consistently enhances retrieval performance while significantly reducing the input context size, which demonstrates the ability of our CoLoR to generalize across diverse retrieval tasks and datasets. We further conduct experiments by training CoLoR on a single domain (i.e., datasets from the same retrieval category) and evaluating its performance on datasets from other domains (other retrieval categories). As shown in Table~\ref{tab:ood_results2}, the results demonstrate that our model generalizes (even in this challenging setting of) across different domains.

\begin{table}[t]
\small
\centering
\caption{\textbf{Results on more challenging out-of-domain settings}, where models are trained on the datasets from the single domain and then evaluated on the datasets from other domains.}
\vspace{-0.05in}
\label{tab:ood_results2}
\resizebox{0.475\textwidth}{!}{
\begin{tabular}{llcccccc}
\toprule
    &   &  \multicolumn{2}{c}{Raw Passage} &  \multicolumn{2}{c}{Comp w/ Phi} &  \multicolumn{2}{c}{CoLoR*} \\

 \cmidrule(l{2pt}r{2pt}){3-4}  \cmidrule(l{2pt}r{2pt}){5-6}  \cmidrule(l{2pt}r{2pt}){7-8} 
 \textbf{Training} & \textbf{Evaluation}   & Perf. &  Comp.  & Perf. &Comp.& Perf.  &Comp. \\

 \midrule
 \midrule
 
Fact- & \textit{Multi-document} \\
checking & HotPotQA  & 0.85 & 1.00x & 0.85 & 1.21x & 0.87   & 1.36x \\
 & MuSiQue & 0.35 & 1.00x & 0.41 & 1.39x & 0.40   & 1.54x \\
 & QAMPARI & 0.56 & 1.00x & 0.55 & 1.33x & 0.56   & 1.48x \\
 & QUEST   & 0.33 & 1.00x & 0.31 & 2.03x & 0.32   & 2.33x \\
  \noalign{\vskip 0.25ex}\cdashline{2-8}\noalign{\vskip 0.75ex}
 & \textit{Argument} \\
 & ArguAna     & 0.28 & 1.00x & 0.27 & 2.26x & 0.33   & 2.72x \\
 & Touché-2020  & 0.76 & 1.00x & 0.79 & 3.79x & 0.82   & 4.56x \\
 \noalign{\vskip 0.25ex}\cdashline{2-8}\noalign{\vskip 0.75ex}
& Average   & 0.52 & 1.00x & 0.53 & 2.00x & 0.55   & 2.33x \\
\midrule
 Multi- & \textit{Fact-checking} \\
 document & FEVER   & 0.95 & 1.00x & 0.95 & 1.83 & 0.95   & 2.14 \\
 & SciFact  & 0.71 & 1.00x & 0.71 & 1.78x & 0.73   & 2.13x \\
 \noalign{\vskip 0.25ex}\cdashline{2-8}\noalign{\vskip 0.75ex}
 & \textit{Argument} \\
 & ArguAna    & 0.28 & 1.00x & 0.27 & 2.26x & 0.33 & 2.73x \\
 & Touché-2020  & 0.76 & 1.00x & 0.79 & 3.79x & 0.82   & 4.65x  \\
 \noalign{\vskip 0.25ex}\cdashline{2-8}\noalign{\vskip 0.75ex}
 & Average  & 0.61 & 1.00x & 0.61 & 2.34x & 0.63   & 2.81x \\

\bottomrule

\end{tabular}
}
\end{table}
\begin{figure}[t!]
    \centering
    \includegraphics[width=0.775\linewidth]
    {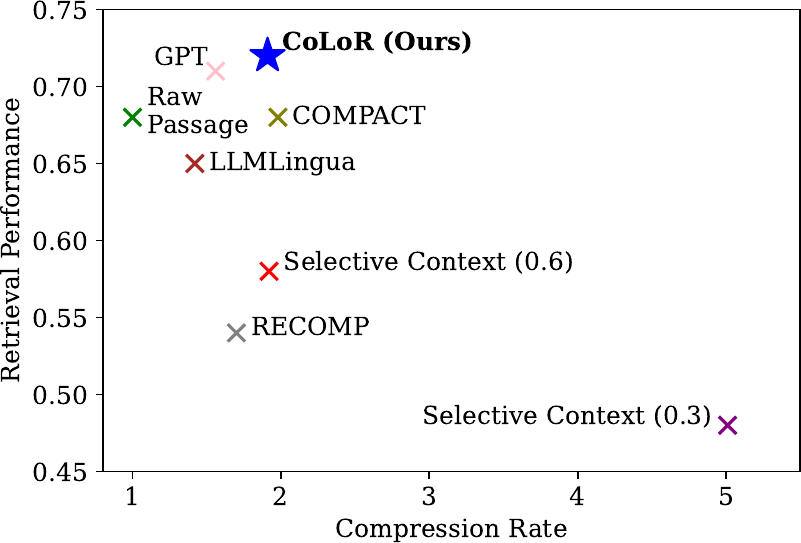}
    \vspace{-0.05in}
    \caption{\textbf{The trade-off of different methods}, showing their compression rate (x-axis) and retrieval performance (y-axis).}
    \label{fig:balance}
    \vspace{-0.05in}
\end{figure}

\paragraph{Trade Off Between Compression Rate and Retrieval Performance}
To examine the trade-off between the compression rate and retrieval performance, we visualize and analyze them in Figure~\ref{fig:balance}. First of all, we observe that, while extremely high compression rates (such as those achieved by using the compression ratio of 0.3 with the Selective Context baseline) drastically reduce the input size, they also lead to significant information loss (potentially due to the removal of crucial information for retrieval), resulting in the retrieval performance drop of 20 on average compared to using the raw passages. This observation highlights the critical trade-off between compression and performance: simply maximizing compression for efficiency compromises accuracy. In contrast, the proposed CoLoR effectively balances this trade-off, ensuring that the reduction in context size does not sacrifice critical information, thanks to our training strategy that guides the model to prefer the compressed passages of successful retrieval over the ones with unsuccessful retrieval (while enforcing brevity as well). 

\begin{table}[t]
\caption{\textbf{Analysis on computational costs of compression strategies} across different methods on the FIQA dataset.}
\vspace{-0.05in}
\label{tab:compression}
\small
\centering
\resizebox{0.475\textwidth}{!}{
\renewcommand{\arraystretch}{1.0}
\begin{tabular}{llcccc}
\toprule
 Methods  & Base Models & \makecell{\# of \\ Params.} & \makecell{Time \\ (Secs)} & R@1 & Comp. \\
\midrule
 Comp. w/ Phi & Phi3 mini  & 3.8B & 1485.63  & 0.68 & 2.31x \\
 COMPACT & Mistral 7B & 7.3B & 3473.95 & 0.48 & 2.56x \\
 \noalign{\vskip 0.25ex}\cdashline{1-6}\noalign{\vskip 0.75ex}
 CoLoR (Ours) & Phi3 mini  & 3.8B & 1290.82  & 0.73 & 2.82x \\

\bottomrule

\end{tabular}
}
\end{table}

\begin{figure}[t!]
    \centering
    \includegraphics[width=0.775\linewidth]{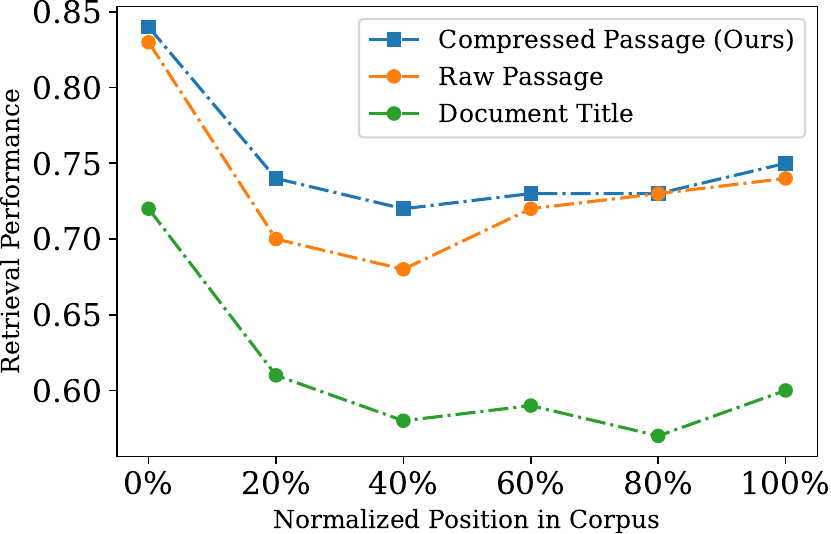}
    \vspace{-0.05in}
    \caption{\textbf{Results with varying the position of (compressed) passages} associated with the query within the corpus, where 0\% (on the x-axis) represents beginning.}
    \label{fig:positional}
    \vspace{-0.075in}
\end{figure}

\begin{table}[t]
    \small
    \centering
    \caption{\textbf{Results with varying the base LMs for CoLoR}, namely Phi3, Mistral, and Llama. Note that Average indicates the average performance across all 9 datasets. Please refer to Table~\ref{tab:full_model_results} for results on other datasets.}
    \vspace{-0.05in}
    \label{tab:model_results}
    \resizebox{0.475\textwidth}{!}{
    \renewcommand{\arraystretch}{1.0}
    \renewcommand{\tabcolsep}{1.0mm}
    \begin{tabular}{lccccccccc}
    \toprule
      & \multicolumn{2}{c}{FIQA} &  \multicolumn{2}{c}{SciFact} &  \multicolumn{2}{c}{MuSiQue} & \multicolumn{2}{c}{Average} \\
    
     \cmidrule(l{2pt}r{2pt}){2-3} \cmidrule(l{2pt}r{2pt}){4-5}  \cmidrule(l{2pt}r{2pt}){6-7} \cmidrule(l{2pt}r{2pt}){8-9} 
     & R@1 & Comp. & R@1 & Comp. & F1@5 & Comp. & Perf. & Comp. \\
     \midrule
     \midrule
     
    Phi3 mini & 0.68 & 2.31x &  0.71 & 1.78x & 0.41 &  1.39x & 0.68 & 1.63x \\
    + CoLoR & 0.67 & 2.82x & 0.75 & 2.12x & 0.42 & 1.55x & 0.72 & 1.91x  \\
     \noalign{\vskip 0.25ex}\cdashline{1-9}\noalign{\vskip 0.75ex}
    
     Mistral 7B & 0.60 & 1.55x & 0.73 & 1.46x & 0.39 & 1.09x & 0.69 & 1.21x \\
    + CoLoR   & 0.63 & 3.07x  & 0.80 & 3.03x & 0.40 & 1.71x & 0.71 & 2.25x \\
     \noalign{\vskip 0.25ex}\cdashline{1-9}\noalign{\vskip 0.75ex}
    
     Llama 3.2 & 0.58 & 2.19x & 0.70 & 2.17x & 0.39 & 1.63x & 0.69 & 1.84x \\
     + CoLoR & 0.61 & 2.83x & 0.72 & 3.03x & 0.39 & 1.71x & 0.69 & 2.15x \\
    
    \bottomrule
    
    \end{tabular}
    }
\end{table}

\begin{table}[t]
\caption{\textbf{Results of an ablation study}, where SFT refers to supervised fine-tuning, and ORPO w/ Reg refers to our full CoLoR model with the dynamic regularization term. Average indicates the average performance across all 9 datasets.}
\vspace{-0.05in}
\label{tab:ablation_results}
\small
\centering
\resizebox{0.475\textwidth}{!}{
\renewcommand{\arraystretch}{1.125}
\begin{tabular}{lccccccccc}
\toprule
  & \multicolumn{2}{c}{MS MARCO} &  \multicolumn{2}{c}{MuSiQue} & \multicolumn{2}{c}{QUEST} &  \multicolumn{2}{c}{Average} \\

 \cmidrule(l{2pt}r{2pt}){2-3} \cmidrule(l{2pt}r{2pt}){4-5}  \cmidrule(l{2pt}r{2pt}){6-7} \cmidrule(l{2pt}r{2pt}){8-9} 
 & R@1 & Comp. & F1@5 & Comp. &  F1@3 & Comp. &Perf. & Comp. \\
 \midrule
 \midrule
 Base Model & 0.92 & 1.41x & 0.41 & 1.39x & 0.31 & 2.03x & 0.71 & 1.63x  \\
 + SFT  & 0.88 & 1.42x & 0.39 & 1.39x & 0.32 & 2.18x & 0.72 & 1.65x  \\
 + ORPO & 0.94 & 1.59x & 0.40 & 1.53x & 0.31 & 2.31x & 0.71 & 1.86x \\
\noalign{\vskip 0.25ex}\cdashline{1-9}\noalign{\vskip 0.75ex}
 + ORPO w/ Reg. & 0.95 & 1.41x & 0.42 & 1.55x & 0.33 & 2.39x & 0.72 & 1.91x \\

\bottomrule

\end{tabular}
}
\vspace{-0.05in}
\end{table}

 \begin{table*}[t]
\caption{\textbf{Results of different retrieval approaches with raw and compressed passages}. In the first couple of columns, Types refers to retrieval types, and Formats refers to corpus formats. Average indicates the performance over all 9 datasets.}
\vspace{-0.05in}
\label{tab:retriever_results}
\small
\centering
\resizebox{0.975\textwidth}{!}{
\renewcommand{\tabcolsep}{3.0mm}
\begin{tabular}{llccccccccc}
\toprule
  & &  \multicolumn{2}{c}{FEVER} &  \multicolumn{2}{c}{MS MARCO} &  \multicolumn{2}{c}{HotpotQA} & \multicolumn{2}{c}{Average} \\

 \cmidrule(l{2pt}r{2pt}){3-4} \cmidrule(l{2pt}r{2pt}){5-6}  \cmidrule(l{2pt}r{2pt}){7-8} \cmidrule(l{2pt}r{2pt}){9-10} 
 \textbf{Types} & \textbf{Formats} & R@1 & Comp. & R@1 & Comp. & F1@2 & Comp. & Perf. & Comp. \\
 \midrule
 \midrule
BM25 & Raw Passage & 0.93 & 1.00x & 0.78 & 1.00x & 0.82 & 1.00x & 0.67 & 1.00x \\
 & Comp. w/ GPT & 0.91 & 1.74x & 0.78 & 1.39x & 0.80 & 1.20x & 0.64 & 1.56x \\
 & CoLoR (Ours) & 0.91 & 2.15x & 0.70 & 1.63x & 0.77 & 1.37x & 0.62 & 1.91x \\
\noalign{\vskip 0.25ex}\cdashline{1-10}\noalign{\vskip 0.75ex}

DPR & Raw Passage & 0.89 & 1.00x & 0.85 & 1.00x & 0.79 & 1.00x & 0.61 & 1.00x  \\
  & Comp. w/ GPT  & 0.89 & 1.74x  & 0.85 & 1.39x & 0.77 & 1.20x & 0.62 & 1.56x \\
 & CoLoR (Ours) & 0.91 & 2.15x  & 0.88 & 1.63x & 0.81 & 1.37x & 0.63 & 1.91x \\

\noalign{\vskip 0.25ex}\cdashline{1-10}\noalign{\vskip 0.75ex}

LCLM & CoLoR (Ours) & 0.94 & 2.15x & 0.95 & 1.63x & 0.86 & 1.37x &  0.72 & 1.91x \\

\bottomrule

\end{tabular}
}

\end{table*}

\paragraph{Analysis on Compression Process}
In Table~\ref{tab:compression}, on the same hardware constraint, we find that our CoLoR is faster in compressing the full corpus of the FIQA dataset (which takes around 20 minutes) than other compression approaches, but also, with the compressed passages from CoLoR, we achieve the superior retrieval performance and compression ratio. Specifically, COMPACT has a longer compression time, as it adopts an iterative approach to compress the text. In contrast, the compression process with CoLoR requires minimal computational burdens. We also note that the passage compression is a highly efficient process in terms of LCLM retrieval, as it can be performed only once, and its outputs are reused (and cached) in inference.

\paragraph{Analysis on Passage Position}
In Figure~\ref{fig:positional}, we analyze the position of passages within the input context (associated with the query in the same context), to see the potential lost-in-the-middle problem~\cite{Liu2023LostIT} in the context of LCLM retrieval: the retrieval performance can be decreased if the relevant passages to the query are placed in the middle of the input sequence. To measure this, we place documents at intervals of 0\%, 20\%, 40\%, 60\%, 80\%, and 100\% across the input corpus, and compare three different approaches: Raw Passage, Document Title, and our CoLoR. Then, similar to the finding in~\citet{Liu2023LostIT}, placing documents towards the middle leads to a certain level of performance degradation across all methods. Yet, interestingly, our proposed compression method (CoLoR) mitigates the lost-in-the-middle issue, since it not only filters out irrelevant information within passages during compression but also allows for a more compact use of the input context. 

\paragraph{CoLoR with Different LMs}
To see whether the proposed CoLoR is versatile across different underlying LMs in generating compressed passages, we vary them with three different LMs: Phi-3-mini-4k-instruct, Mistral-7B-Instruct-v0.3, and Llama-3.2-3B-Instruct. Then, as shown in Table~\ref{tab:model_results}, we observe the consistent improvements of our CoLoR in both compression rate and retrieval performance across all models. This demonstrates that CoLoR and its training methodology is not limited to a specific model but can be effectively generalized to others. We provide results with all datasets in Table~\ref{tab:full_model_results}.

\paragraph{Ablation Study}
To see the effectiveness of each component of our CoLoR, we perform an ablation study and present the results in Table~\ref{tab:ablation_results}. First of all, we observe that, while Supervised Fine-Tuning (SFT) yields strong retrieval performance, the compression rate remains comparable to the untrained method. On the other hand, by utilizing preference optimization with Odds Ratio Preference Optimization (ORPO), we observe an improved compression rate, though this comes with a slight performance degradation. However, the proposed dynamic regularization term (for compressed passage length) mitigates this trade-off, further improving both the compression ratio and retrieval performance, reaffirming the overall efficacy of our proposed CoLoR approach in both the efficiency and effectiveness. More detailed results are in Table~\ref{tab:full_ablation_results} of Appendix.

\paragraph{Adaptation of CoLoR to Conventional Retrieval}
In Table~\ref{tab:retriever_results}, we investigate how using compressed passages impacts the performance of conventional sparse and dense retrievers, as they can bring an additional benefit of faster indexing thanks to the reduced passage length. First, for the sparse retriever (BM25), performance tends to decrease when using compressed passages, likely due to the loss of lexical information that BM25 relies on to match documents based on exact lexical similarities. In contrast, the dense retriever (DPR) shows performance improvements with compressed passages. We conjecture that this might be because the underlying LM for dense retrieval already contains much of the passage's information within its parameters, and, as a result, compressing the passage still retains essential details in making valuable representations for it while additionally filtering out irrelevant content (that might lead to noise in embedding). However, despite these gains in dense retrieval with the proposed CoLoR, the performance of LCLM retrieval coupled with CoLoR is substantially better than conventional retrieval methods.

 \begin{table}[t]
\caption{\textbf{Manual evaluation results} on three sampled passages per dataset. We report the average number of total facts (Facts), query-supportive facts (Sup. Facts), the proportion of supportive facts to total facts (Ratio), and the token count.}
\vspace{-0.05in}
\label{tab:humaneval_result}
\small
\centering
\resizebox{0.475\textwidth}{!}{
\renewcommand{\arraystretch}{1.075}
\begin{tabular}{lcccc}
\toprule
 Methods & Facts& Sup. Facts & Ratio & Tokens \\
 \midrule
 Raw Passages & 14.13 & 1.91 & 13.52 & 210.93 \\
 CoLoR (Ours) & 9.26 & 1.74 & 18.80 & 93.41 \\

\bottomrule

\end{tabular}
}
\vspace{-0.05in}
\end{table}

\paragraph{Qualitative Analysis with Manual Evaluation}
To see whether query-relevant information is preserved after passage compression, we manually compare atomic facts in the compressed passages to ones in the raw passages, with randomly sampled three examples from each of all datasets with two individual annotators. As shown in Table~\ref{tab:humaneval_result}, while the total number of facts in the compressed passages decreases, the number of query-relevant facts is only slightly reduced (from 1.91 to 1.74 per passage on average). Also, when we look at the proportion of relevant facts to total facts (Ratio), this proportion increases, indicating that the compressed passages contain a higher density of query-related atomic facts (while the proportion of noisy, query-irrelevant information is reduced), which may support the performance improvement of our CoLoR. Additionally, we provide the case study on the compressed passages in Figure~\ref{tab:case_study}.

\section{Conclusion}
In this work, we introduced \textbf{Co}mpression for \textbf{Lo}ng Context Langauge Model \textbf{R}etrieval (\textbf{\ours}), a method specifically designed to improve efficiency and effectiveness of LCLM retrieval by transitioning from raw to compressed passages. Specifically, the proposed CoLoR, trained with the synthesized preference data (based on retrieval outcomes of the compressed passages) and regularization loss for their lengths, optimizes both brevity and retrieval performance. Through our extensive experiments conducted across 9 datasets spanning single- and multi-document retrieval tasks, we demonstrated that CoLoR not only achieves a 6\% improvement in retrieval performance but also reduces context size by a factor of 1.91 over the standard LCLM retrieval, which further surpasses existing text compression methods. These highlight the significant advantage of compressed passages to enhance efficiency for LCLM retrieval by reducing the computational load and its associated costs, all while even improving retrieval accuracy, making it more scalable and practical for real-world applications.

\section*{Limitations}
While our proposed CoLoR approach demonstrates significant advantages in LCLM retrieval, there are still areas that future work may explore. First, following the LCLM retrieval benchmark setup~\cite{Lee2024LOFT}, our experiments are conducted with a maximum context length of 128K tokens, and, while this context length is indeed very large and it has been increasingly extended further, in real-world applications, the size of the corpus can be much larger (even after utilizing our compression method), which may necessitate further modifications of the overall LCLM retrieval framework. Yet, developing the new process for LCLM retrieval is beyond the scope of our work and we leave it as future work. Another consideration is the compression process: it introduces an additional step before retrieval; however, this is not a big deal as it only needs to be performed once as like the indexing process of sparse and dense retrieval approaches.

\section*{Ethics Statement}
It is worth noting that, similar to any other retrieval approaches, the retrieval corpus may contain harmful or offensive content, and the compressed passages could potentially reflect these biases. Also, additional biases may be induced during the training process of LCLMs. Although addressing these concerns are obviously beyond the scope of our work, we acknowledge the importance of implementing the safeguards in future research to ensure that the retrieval process remains safe and fair.

\section*{Acknowledgements}
This work was supported by the National Research Foundation of Korea (NRF) grant funded by the Korea government (MSIT) (No. RS-2023-00256259), the grant of the Korea Machine Learning Ledger Orchestration for Drug Discovery Project (KMELLODDY) funded by the Ministry of Health \& Welfare and Ministry of Science and ICT, Republic of Korea (grant number: RS-2024-12345678), the Artificial intelligence industrial convergence cluster development project funded by the Ministry of Science and ICT (MSIT, Korea) \& Gwangju Metropolitan City, and the Institute for Information \& communications Technology Planning \& Evaluation (IITP) grant funded by the Korea government (MSIT) (RS-2019-II190075, Artificial Intelligence Graduate School Program (KAIST), and No. RS-2022-II220713, Meta-learning Applicable to Real-world Problems).

\bibliography{custom}

\appendix

\clearpage

 \begin{table}[t]
\caption{Statistics of the data samples generated for preference optimization for training CoLoR, which includes the number of samples per dataset, and the average length of reject and chosen tokens. \dag~denotes multi-document retrieval datasets.}
\vspace{-0.1in}
\label{tab:dataset_stat}
\small
\centering
\resizebox{0.475\textwidth}{!}{
\renewcommand{\arraystretch}{1.075}
\begin{tabular}{lccc}
\toprule
Dataset & \# of Samples & Avg. Rejected Token & Avg. Chosen Token \\
 \midrule
 FEVER & 483 & 236.19 & 152.63 \\
 FIQA & 455 & 148.15 & 88.22 \\
 MS MARCO & 1061 & 76.06 & 51.3 \\
 NQ & 635 & 158.16 & 111.67 \\
 SciFact & 198 & 229.84 & 156.26 \\
 HotPotQA\textsuperscript{\dag} & 544 & 93.52 & 71.11 \\
 MuSiQue\textsuperscript{\dag} & 12 & 107.58 & 80.5 \\
 QAMPARI\textsuperscript{\dag} & 30 & 107.23 & 82.77 \\
 \noalign{\vskip 0.25ex}\cdashline{1-4}\noalign{\vskip 0.75ex}
 \textbf{Total} & 3418 & 135.61 & 91.36 \\

\bottomrule

\end{tabular}
}

\end{table}

 \begin{table}[t]
\caption{Statistics of the benchmark retrieval datasets for experiments. \dag~denotes multi-document retrieval datasets.}
\vspace{-0.1in}
\label{tab:dataset_stat2}
\small
\centering
\resizebox{0.475\textwidth}{!}{
\renewcommand{\arraystretch}{1.075}
\begin{tabular}{lccccc}
\toprule
 Dataset& \# of Passage & \makecell{Avg. \\ Passage Token }& \makecell{Avg. Comp. \\ Passage Token } & Comp. Ratio \\
\midrule
 FEVER & 588& 169  & 78.60 & 2.15 \\
 FIQA& 531& 190  & 67.43 & 2.82 \\
 MS MARCO & 1,174& 76   & 46.76& 1.63 \\
 NQ& 883& 104  & 64.07 & 1.62 \\
 SciFact& 357& 291  & 137.03& 2.12& \\ 
 HotPotQA\textsuperscript{\dag} & 1,222& 69   & 50.43& 1.37 \\
 MuSiQue\textsuperscript{\dag} & 824& 112  & 72.22 & 1.55 \\
 QAMPARI\textsuperscript{\dag} & 755& 125  & 83.26 & 1.50 \\
 QUEST\textsuperscript{\dag} & 328& 325  & 135.77& 2.39 \\
 \noalign{\vskip 0.25ex}\cdashline{1-6}\noalign{\vskip 0.75ex}
 Average & 740.22& 162.33  & 81.73  & 1.91 \\

\bottomrule

\end{tabular}
}

\vspace{-0.05in}
\end{table}

\section{Additional Experimental Setups}
\label{appendix:setups}

\paragraph{Data Collection Details}
We collect compressed versions of passages to train CoLoR. The diversity of the compressed passages (including good and bad ones) is crucial to construct data for preference optimization, and to ensure this, we prompt different LLMs, such as Phi-3-mini-4k-instruct~\cite{phi3}, Mistral-7B-Instruct-v0.3~\cite{mistral}, and Llama-3.2-3B-Instruct~\cite{llama3}, and GPT-4o-mini~\cite{GPT4}, with the same prompt: \texttt{Summarize the following content: \{passage\}}.

\paragraph{Fine-tuning Details}
For Supervised Fine-Tuning (SFT), we use a learning rate of 5e-6. Similarly, for ORPO with Phi and Llama, we use $\lambda$ of 2.5 and a learning rate of 1e-6, while, for Mistral, we change the learning rate to 5e-6. Also, all models are trained for 10 epochs with a batch size of 8, and the best epoch is selected based on the validation set. Lastly, we use the TRL library\footnote{https://github.com/huggingface/trl} for training.

\paragraph{Computational Resources}
We perform training and inference on all baselines and our model by using one of the NVIDIA RTX A6000 and NVIDIA RTX A5000 GPUs, depending on their availability. With those GPUs, the time required to train our model over 10 epochs ranges from 3 to 5 hours.

\paragraph{Deep Learning Libraries}
In our experiments, we utilize the following deep learning libraries: PyTorch~\citep{pytorch}, Transformers~\citep{hf_transformers}, SentenceTransformers~\citep{SentenceBERT}, and BEIR~\citep{beir}. Also, BM25 is implemented using a python library rank\_bm25\footnote{https://github.com/dorianbrown/rank\_bm25}, while, for DPR, we use a BEIR framework\footnote{https://github.com/beir-cellar/beir}. Other baselines are sourced from publicly available checkpoints on their repositories\footnote{https://huggingface.co/cwyoon99/CompAct-7b}\footnote{https://github.com/liyucheng09/Selective\_Context}. 

\paragraph{Datasets Details}

In Table~\ref{tab:dataset_stat}, we provide the statistics of the data samples that we create for training our compression model (CoLoR). Note that, among all samples, we randomly use 3,077 samples for training and 341 samples for validation. Also, Table~\ref{tab:dataset_stat2} summarizes the retrieval datasets. We follow the experimental setup from the LOFT benchmark~\cite{Lee2024LOFT}, ensuring that the number of passages included in the LCLM context matches those in~\citet{Lee2024LOFT}. For all experiments, we use GPT-4o-mini as the underlying LCLM, which supports a context length of 128K tokens.

\section{Additional Experimental Results}
\label{appendix:results}
 \begin{table}[t]
\caption{Results on the long context retrieval benchmark. }
\vspace{-0.1in}
\label{tab:long_results}
\small
\centering
\resizebox{0.475\textwidth}{!}{
\renewcommand{\arraystretch}{1.0}
\renewcommand{\tabcolsep}{1.0mm}
\begin{tabular}{lccccccccc}
\toprule
  & \multicolumn{2}{c}{NQ} &  \multicolumn{2}{c}{2WikimQA} &  \multicolumn{2}{c}{NarrativeQA} & \multicolumn{2}{c}{Average} \\

 \cmidrule(l{2pt}r{2pt}){2-3} \cmidrule(l{2pt}r{2pt}){4-5}  \cmidrule(l{2pt}r{2pt}){6-7} \cmidrule(l{2pt}r{2pt}){8-9} 
 & R@1 & Comp. & R@1 & Comp. & R@1 & Comp. & R@1 & Comp. \\
 \midrule
 \midrule
 
 Raw Passage & 0.92 & 1.00x & 0.42 & 1.00x & 0.22 & 1.00x & 0.52 & 1.00x \\

 Compressed Passage & 0.95 & 63.02x & 0.72 & 29.64x & 0.56 & 555.93x & 0.74 & 216.2x \\
  
\bottomrule

\end{tabular}
}

\vspace{-0.05in}
\end{table}

 \begin{table*}[t]
\caption{\textbf{Full results with all datasets by varying the base LM for CoLoR}. \dag~indicates multi-document retrieval datasets, and * denotes out-of-domain datasets (that are not used for training CoLoR).}
\vspace{-0.1in}
\label{tab:full_model_results}
\small
\centering
\resizebox{0.875\textwidth}{!}{
\renewcommand{\arraystretch}{1.075}
\begin{tabular}{lcccccccccc}
\toprule
& \multicolumn{2}{c}{FEVER} & \multicolumn{2}{c}{FIQA} & \multicolumn{2}{c}{MS MARCO} & \multicolumn{2}{c}{NQ} & \multicolumn{2}{c}{SciFact} \\

\cmidrule(l{2pt}r{2pt}){2-3} \cmidrule(l{2pt}r{2pt}){4-5} \cmidrule(l{2pt}r{2pt}){6-7} \cmidrule(l{2pt}r{2pt}){8-9} \cmidrule(l{2pt}r{2pt}){10-11}
 \textbf{Methods} & R@1 & Comp. & R@1 & Comp. & R@1 & Comp.& R@1 & Comp.& R@1 & Comp. \\

 \midrule
 \midrule
 
 Phi3 mini & 0.95 & 1.83x & 0.68 & 2.31x & 0.92 & 1.41x & 0.99 & 1.39x & 0.71 & 1.78x \\
 + CoLoR & 0.94 & 2.15x & 0.73 & 2.82x & 0.95 & 1.63x & 0.98 & 1.62x & 0.75 & 2.12x \\

\noalign{\vskip 0.25ex}\cdashline{1-11}\noalign{\vskip 0.75ex}
Mistral 7B & 0.96 & 1.36x & 0.58 & 1.55x & 0.90 & 0.98x & 0.98 & 1.01x & 0.70 & 1.46x \\
+ CoLoR & 0.96 & 2.51x & 0.63 & 3.07x & 0.91 & 1.66x & 0.98 & 1.78x & 0.80 & 3.03x \\
 
\noalign{\vskip 0.25ex}\cdashline{1-11}\noalign{\vskip 0.75ex}
  
 Llama 3.2 & 0.96 & 2.03x & 0.35 & 2.19x & 0.52 & 1.47x & 0.95 & 1.68x & 0.71 & 2.17x \\
 + CoLoR & 0.95 & 2.30x & 0.61 & 2.83x & 0.86 & 2.76x & 0.99 & 1.91x & 0.72 & 2.64x \\
\midrule

\midrule

 & \multicolumn{2}{c}{HotPotQA\textsuperscript{\dag}} & \multicolumn{2}{c}{MuSiQue\textsuperscript{\dag}} & \multicolumn{2}{c}{QAMPARI\textsuperscript{\dag}} & \multicolumn{2}{c}{QUEST\textsuperscript{\dag*}} & \multicolumn{2}{c}{Average} \\
 
\cmidrule(l{2pt}r{2pt}){2-3} \cmidrule(l{2pt}r{2pt}){4-5} \cmidrule(l{2pt}r{2pt}){6-7} \cmidrule(l{2pt}r{2pt}){8-9} \cmidrule(l{2pt}r{2pt}){10-11}

 \textbf{Methods} & F1@2 & Comp. & F1@5 & Comp. & F1@5 & Comp.& F1@3 & Comp.& Perf. & Comp. \\
\midrule
 \midrule
 
 Phi3 mini & 0.85 & 1.21x & 0.41 & 1.39x & 0.55 & 1.33x & 0.31 & 2.03x & 0.68 & 1.63x \\
 + CoLoR & 0.86 & 1.37x & 0.42 & 1.55x & 0.55 & 1.50x & 0.33 & 2.39x & 0.72 & 1.91x \\
 
 \noalign{\vskip 0.25ex}\cdashline{1-11}\noalign{\vskip 0.75ex}
 
 Mistral 7B & 0.83 & 0.92x & 0.39 & 1.09x & 0.55 & 1.07x & 0.32 & 1.46x & 0.69 & 1.21x \\
 + CoLoR & 0.85 & 1.43x & 0.40 & 1.71x & 0.56 & 1.68x & 0.33 & 3.35x & 0.71 & 2.25x  \\
 
\noalign{\vskip 0.25ex}\cdashline{1-11}\noalign{\vskip 0.75ex}
 
 Llama 3.2 & 0.84 & 1.42x & 0.39 & 1.63x & 0.52 & 1.64x & 0.31 & 2.33x & 0.69 & 1.84x \\
 + CoLoR & 0.85 & 1.55x & 0.40 & 1.84x & 0.54 & 1.83x & 0.30 & 2.68x & 0.69 & 2.15x \\ 
\bottomrule

\end{tabular}
}

\end{table*}

 \begin{table*}[t]
\caption{\textbf{Full results of the ablation study with all datasets.}  \dag~denotes multi-document retrieval datasets, and * indicates out-of-domain datasets (not used for training CoLoR). SFT refers to supervised fine-tuning, and ORPO w/ Reg denotes CoLoR.}
\vspace{-0.1in}
\label{tab:full_ablation_results}
\small
\centering
\resizebox{0.875\textwidth}{!}{
\renewcommand{\arraystretch}{1.075}
\begin{tabular}{lcccccccccc}
\toprule
& \multicolumn{2}{c}{FEVER} & \multicolumn{2}{c}{FIQA} & \multicolumn{2}{c}{MS MARCO} & \multicolumn{2}{c}{NQ} & \multicolumn{2}{c}{SciFact} \\

\cmidrule(l{2pt}r{2pt}){2-3} \cmidrule(l{2pt}r{2pt}){4-5} \cmidrule(l{2pt}r{2pt}){6-7} \cmidrule(l{2pt}r{2pt}){8-9} \cmidrule(l{2pt}r{2pt}){10-11}
 \textbf{Methods} & R@1 & Comp. & R@1 & Comp. & R@1 & Comp.& R@1 & Comp.& R@1 & Comp. \\

 \midrule
 \midrule

 Base Model & 0.95 & 1.83x & 0.68 & 2.31x & 0.92 & 1.41x & 0.99 & 1.39x & 0.71 & 1.78x \\
 + SFT & 0.95 & 1.82x & 0.67 & 2.28x & 0.88 & 1.42x & 0.99 & 1.42x & 0.80 & 1.83x \\
+ ORPO & 0.94 & 2.11x & 0.64 & 2.75x & 0.94 & 1.59x & 0.99 & 1.59x & 0.74 & 2.05x \\
 
\noalign{\vskip 0.25ex}\cdashline{1-11}\noalign{\vskip 0.75ex}

 + ORPO w/ Reg & 0.94 & 2.15x & 0.73 & 2.82x & 0.95 & 1.63x & 0.98 & 1.62x & 0.75 & 2.12x  \\
\midrule

\midrule

 & \multicolumn{2}{c}{HotPotQA\textsuperscript{\dag}} & \multicolumn{2}{c}{MuSiQue\textsuperscript{\dag}} & \multicolumn{2}{c}{QAMPARI\textsuperscript{\dag}} & \multicolumn{2}{c}{QUEST\textsuperscript{\dag*}} & \multicolumn{2}{c}{Average} \\
 
\cmidrule(l{2pt}r{2pt}){2-3} \cmidrule(l{2pt}r{2pt}){4-5} \cmidrule(l{2pt}r{2pt}){6-7} \cmidrule(l{2pt}r{2pt}){8-9} \cmidrule(l{2pt}r{2pt}){10-11}

 \textbf{Methods} & F1@2 & Comp. & F1@5 & Comp. & F1@5 & Comp.& F1@3 & Comp.& Perf. & Comp. \\
\midrule
 \midrule
 
 Base Model & 0.85 & 1.21x & 0.41 & 1.39x & 0.55 & 1.33x & 0.23 & 2.03x & 0.71 & 1.63x \\
 + SFT & 0.88 & 1.22x & 0.39 & 1.39x & 0.57 & 1.33x & 0.32 & 2.18x & 0.72 & 1.65x \\
+ ORPO & 0.85 & 1.35x & 0.40 & 1.53x & 0.54 & 1.48x & 0.31 & 2.31x & 0.71 & 1.86x \\
 
\noalign{\vskip 0.25ex}\cdashline{1-11}\noalign{\vskip 0.75ex}

 + ORPO w/ Reg & 0.86 & 1.37x & 0.42 & 1.55x & 0.55 & 1.50x & 0.33 & 2.39x & 0.72 & 1.91x \\
\bottomrule

\end{tabular}
}

\vspace{-0.05in}
\end{table*}

\paragraph{Results on Long Context Retrieval Benchmark}
We further evaluate our CoLoR on the long context retrieval scenario, including two long context question-answering datasets from the LongEmbed benchmark~\cite{Zhu2024LongEmbedEE} as well as the original corpus for the Natural Questions (NQ) dataset\footnote{https://github.com/google-research-datasets/natural-questions}. To enable comparisons between different methods, the raw passages are truncated (as the context size with original raw passages exceeds its limit) and the compressed passages are generated using GPT-4o-mini (prompted to create summaries under 200 words)\footnote{Due to the excessive length of passages for these datasets, training CoLoR on them is not feasible within our computational resources, and we leave this as a future work.}. Then, as shown in Table~\ref{tab:long_results}, the compression model reduces passage size by 216.2$\times$, while increasing Recall@1 by 42\%, compared to using (truncated) raw passages, which further strengths the effectiveness of our compression paradigm, particularly in handling lengthy passages.

\begin{figure*}[t!]
\centering
    \includegraphics[width=0.7\linewidth]{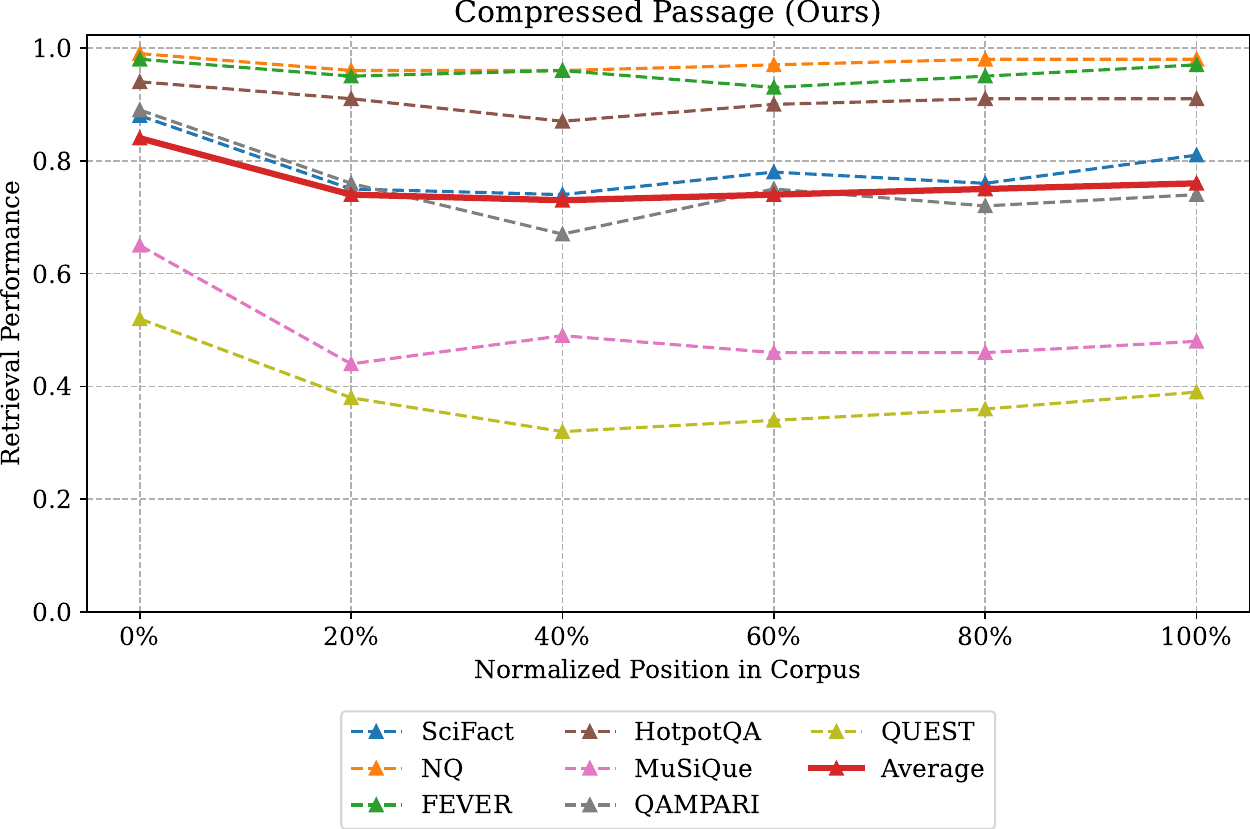}
    \vspace{0.15in}
    
    \includegraphics[width=0.7\linewidth]{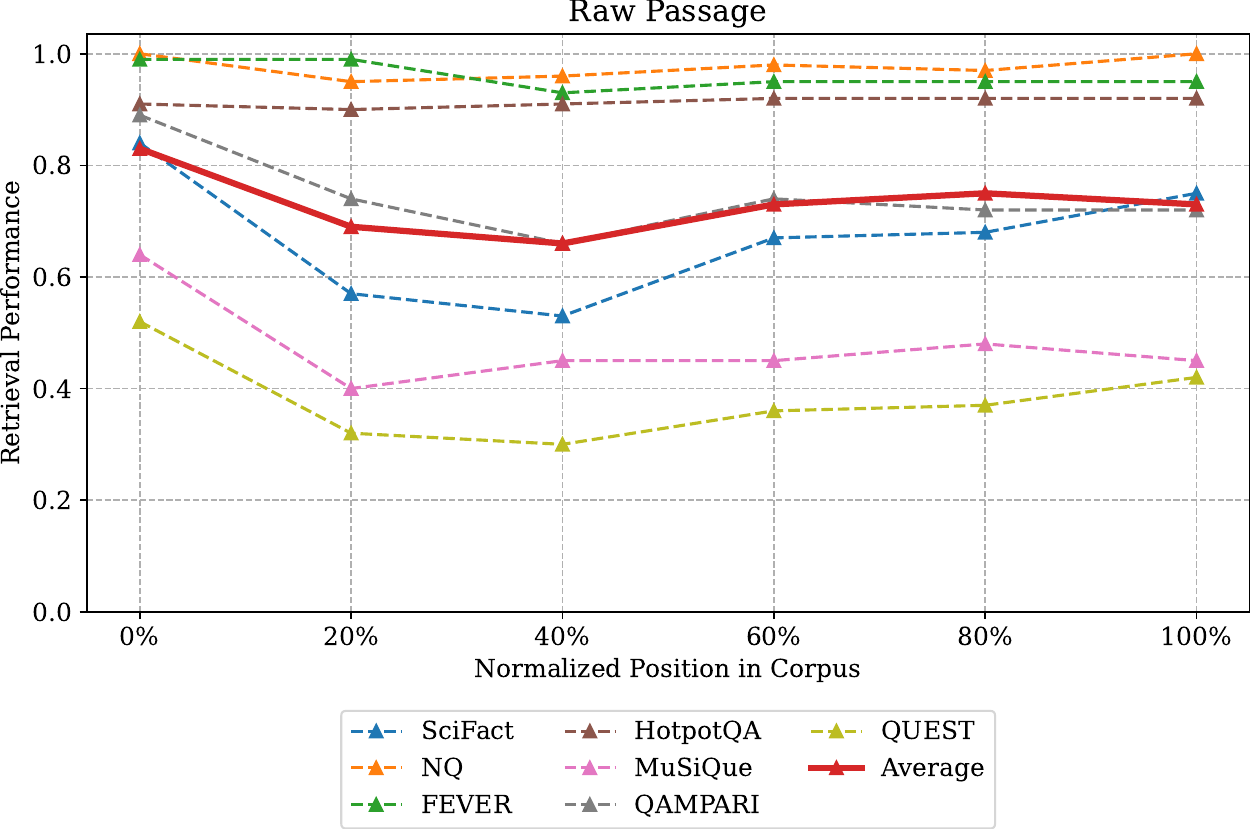}
    \vspace{0.15in}
    
    \includegraphics[width=0.7\linewidth]{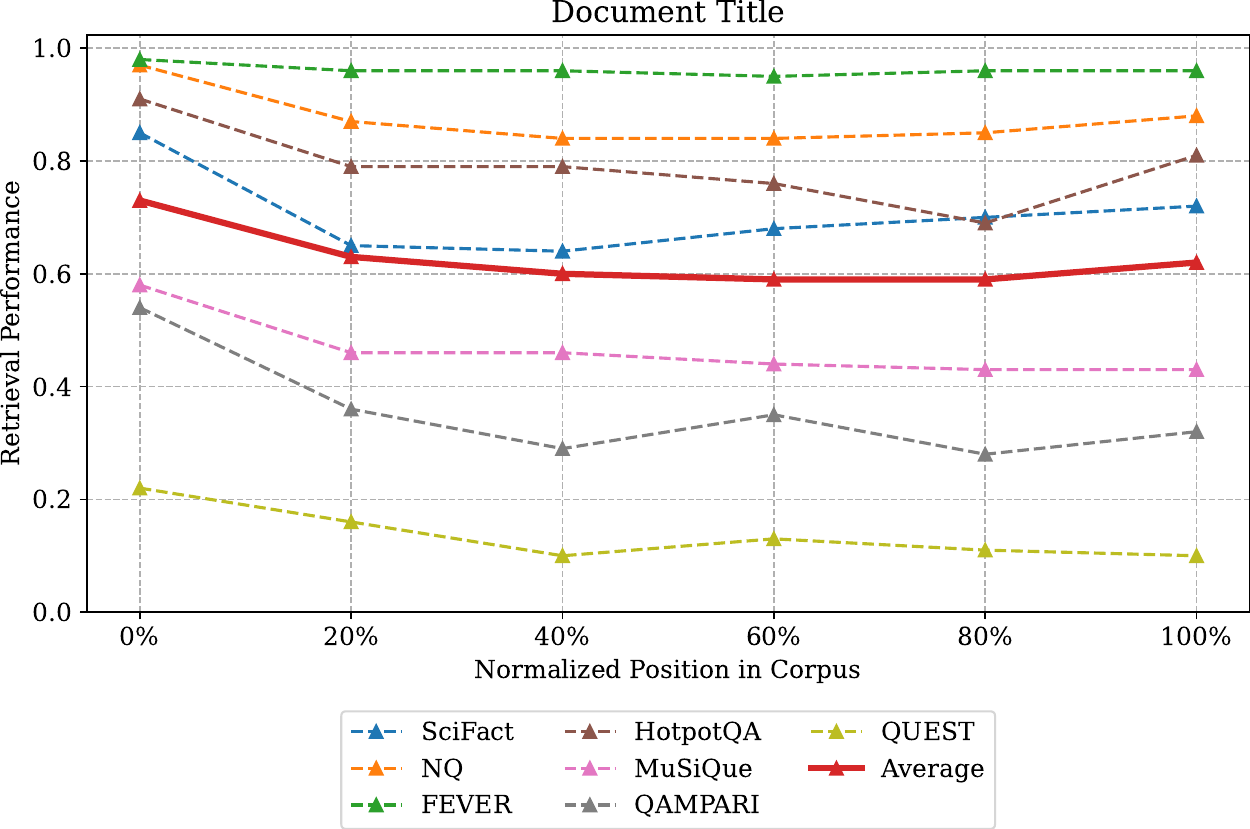}

    \caption{\textbf{Results with varying the position of (compressed) passages for all datasets}. Specifically, we arbitrarily adjust the positions of the gold and few-shot passages within the corpus relative to the query (0\% represents the beginning). The figures at the top, middle, and bottom represent the results with CoLoR, raw passage, and document title, respectively.}
    \label{fig:full_positional}
    
\end{figure*}

\paragraph{Full Results on Analyses} 
In Table~\ref{tab:full_model_results} and Table~\ref{tab:full_ablation_results}, we provide the full results of varying the base LM and the ablation study with all datasets, respectively. Also, we provide the results of analysis on passage position with all datasets in Figure~\ref{fig:full_positional}. 

\paragraph{Case Study}
We provide the case study on the compressed passages generated by different approaches in Table~\ref{tab:case_study}, which shows that the compressed passages from our CoLoR tend to leading to the retrieval success and tend to be shorter. 

\paragraph{Prompt Details} 
For the prompt construction, we follow the Corpus-in-Context prompting approach from prior work~\cite{Lee2024LOFT}. An example prompt for the NQ dataset is provided in Table~\ref{tab:prompt}, and, for more examples and details on the prompt, please refer to~\citet{Lee2024LOFT}.

 \begin{table*}[t!]
\caption{\textbf{Case study} on the retrieval sample from the FIQA dataset.}
\vspace{-0.1in}
\label{tab:case_study}
\centering
\resizebox{0.975\textwidth}{!}{
\begin{tabular}{lcc}
\toprule

\makecell{
 \multicolumn{1}{p{.15\textwidth}}{\textbf{Methods}} \\ \multicolumn{1}{p{.15\textwidth}}{\textbf{(\# of Tokens)}}
 } & \textbf{Passage} & \textbf{Prediction}  \\
 \midrule

 \makecell{
 \multicolumn{1}{p{.15\textwidth}}{Query}
 }
  & \makecell{
 \multicolumn{1}{p{.95\textwidth}}{
 If an index goes up because an underlying company issues more shares, what happens to the ETF}} \\

 \midrule
 \midrule
 
 \makecell{
 \multicolumn{1}{p{.15\textwidth}}{Raw Passage} \\ \multicolumn{1}{p{.15\textwidth}}{(315)}
 }
 & \makecell{
 \multicolumn{1}{p{.95\textwidth}}{If a stock that makes up a big part of the Dow Jones Industrial   Average decided to issue a huge number of additional shares, that will   make the index go up. At least this is what should happen, since an   index is basically a sum of the market cap of the contributing   companies. No, indices can have various weightings. The DJIA is a price-weighted index not market-cap weighted. An alternative weighting besides market-cap and price is equal weighting. From Dow Jones: Dow Jones Industrial Average\u2122. Introduced in May 1896, the index, also   referred to as The Dow\u00ae, is a price-weighted measure of 30 U.S.   blue-chip companies. Thus, I can wonder what in the new shares makes the index go up? If a stock is split, the Dow divisor is adjusted as one could easily see how the current Dow value isn't equal to the sum or the share prices of the members of the index. In other cases, there may be a dilution of earnings but that doesn't necessarily affect the stock price directly as there may be options exercised or secondary offerings made. SO if the index, goes up, will the ETF DIA also go up automatically   although no additional buying has happened in the ETF itself? If the index rises and the ETF doesn't proportionally, then there is an arbitrage opportunity for someone to buy the DIA shares that can be redeemed for the underlying stocks that are worth more in this case. Look at the Creation and Redemption Unit process that exists for ETFs.}} & 325818 \textcolor{red}{\xmark} \\
 \midrule

 \makecell{
 \multicolumn{1}{p{.15\textwidth}}{Comp. w/ GPT} \\ \multicolumn{1}{p{.15\textwidth}}{(185)}
 }
 &
 \makecell{
 \multicolumn{1}{p{.95\textwidth}}{The Dow Jones Industrial Average (DJIA) is a price-weighted index of 30 U.S. blue-chip companies, meaning its value is based on the share prices of its components rather than their market capitalizations. If a major stock in the DJIA issues a large number of new shares, the index should theoretically rise, but adjustments like the Dow divisor may affect this relationship. When a stock splits, for example, the divisor is adjusted to maintain the index's integrity. If the index increases, it raises the question of whether the corresponding ETF (DIA) will also rise automatically, despite no new buying in the ETF. If the ETF's price doesn't align proportionally with the index's increase, it may create an arbitrage opportunity, allowing traders to profit by redeeming DIA shares for the underlying, more valuable stocks. Understanding the creation and redemption unit process is essential for grasping this dynamic.}} & 325818 \textcolor{red}{\xmark} \\
 \noalign{\vskip 0.25ex}\cdashline{1-3}\noalign{\vskip 0.75ex}
 
 \makecell{
 \multicolumn{1}{p{.15\textwidth}}{Comp. w/ Phi} \\ \multicolumn{1}{p{.15\textwidth}}{(138)}
 }
 
 & 
 \makecell{
 \multicolumn{1}{p{.95\textwidth}}{The content discusses how the Dow Jones Industrial Average (DJIA) index, a price-weighted measure of 30 U.S. blue-chip companies, can be affected by the issuance of additional shares by a major company. It explains that the index value may not directly reflect the sum of the share prices due to adjustments in the Dow divisor. The content also addresses the relationship between the DJIA index and the ETF DIA, stating that if the index rises but the ETF doesn't proportionally, there may be an arbitrage opportunity for someone to buy DIA shares and redeem them for the underlying stocks that are worth more. The creation and redemption process of ETFs is briefly mentioned." } }& 325818 \textcolor{red}{\xmark} \\
 
 \noalign{\vskip 0.25ex}\cdashline{1-3}\noalign{\vskip 0.75ex}
  
 \makecell{
 \multicolumn{1}{p{.15\textwidth}}{LLMLingua} \\ \multicolumn{1}{p{.15\textwidth}}{(189)}
 }
 
 & \makecell{
 \multicolumn{1}{p{.95\textwidth}}{a stock that the IndustrialAverage to a of that will  . least what, since is a market the., can have various The index. Aning. From: Average™. in index also as Dowed  ., can what the new shares is the Dow is adjusted as one could see the value't sum the the members of the index. In other cases, there may be a dilution of earnings but that doesn't necessarily affect the stock price directly as there may be options exercised or secondary offerings made. SO if the index, goes up, will the ETF DIA also go up automatically   although no additional buying has happened in the ETF itself? If the index rises and the ETF doesn't proportionally, then there is an arbitrage opportunity for someone to buy the DIA shares that can be redeemed for the underlying stocks that are worth more in this case. Look at the Creation and Redemption Unit process that exists for ETFs.} } & 325818 \textcolor{red}{\xmark} \\
 \noalign{\vskip 0.25ex}\cdashline{1-3}\noalign{\vskip 0.75ex}

\makecell{
 \multicolumn{1}{p{.15\textwidth}}{Selective \;\;\; Context (0.3)} \\ \multicolumn{1}{p{.15\textwidth}}{(69)}
 }

& \makecell{
\multicolumn{1}{p{.95\textwidth}}{of the Dow Jones Industrial Average to of, up. is what happen of of....: Dow Jones Industrial Average in, the to as is of., what up is, isn't to of of., be ofn't necessarily the stock price may be. up, up?n't then is for to buy be for are more. at.}} & 482415 \textcolor{red}{\xmark}  \\
\noalign{\vskip 0.25ex}\cdashline{1-3}\noalign{\vskip 0.75ex}

\makecell{
 \multicolumn{1}{p{.15\textwidth}}{Selective \;\;\; Context (0.6)} \\ \multicolumn{1}{p{.15\textwidth}}{(175)}
 }
 
& \makecell{
\multicolumn{1}{p{.95\textwidth}}{up a big part of the Dow Jones Industrial Average to a huge number of additional shares, that make the index go up. is what happen, is a sum of the market cap of., can. is a price-weighted index weighted. and is equal weighting.: Dow Jones Industrial Average in May, the also referred to as is a price-weighted measure of 30 U.S. blue-chip companies., I can what go up? is, is as see the current Dow valuen't equal to the sum the share prices of the members of the index., there may be a dilution of that doesn't necessarily affect the stock price as there may be or made. up, will go up has happened in the ETF itself? andn't then there is an arbitrage opportunity for to buy be redeemed for the underlying stocks that are more in this case. at for}} & 418150 \textcolor{green}{\ding{51}} \\
\noalign{\vskip 0.25ex}\cdashline{1-3}\noalign{\vskip 0.75ex}
 
\makecell{
 \multicolumn{1}{p{.15\textwidth}}{COMPACT} \\ \multicolumn{1}{p{.15\textwidth}}{(116)}
 }
& \makecell{
\multicolumn{1}{p{.95\textwidth}}{The Dow Jones Industrial Average (DJIA) or "Dow 30" is a price-weighted average of 30 significant stocks traded on the New York Stock Exchange (NYSE) and the Nasdaq. The DJIA is one of the most widely followed equity indices. The DJIA is calculated by summing the individual prices of the 30 stocks in the index, then dividing that total by a divisor. If a stock that makes up a big part of the Dow Jones Industrial Average decided to issue a huge number of additional shares, that will make the index go up.}} & 325818 \textcolor{red}{\xmark}\\
\midrule
 
\makecell{
 \multicolumn{1}{p{.15\textwidth}}{CoLoR (Ours)} \\ \multicolumn{1}{p{.15\textwidth}}{(103)}
 } & \makecell{\multicolumn{1}{p{.95\textwidth}}{A stock issuing additional shares can cause the Dow Jones Industrial Average (DJIA) to rise due to its price-weighted nature. However, the DJIA's value isn't directly equal to the sum of its member stocks' prices. In cases of stock splits or dilution of earnings, the Dow divisor is adjusted. If the DJIA rises and the ETF DIA doesn't proportionally increase, there's an arbitrage opportunity for someone to buy DIA shares and redeem them for the underlying stocks worth more.}} & 418150 \textcolor{green}{\ding{51}} \\

\bottomrule

\end{tabular}
}
\end{table*}

\begin{table*}[t]
    \small
    \centering
    \caption{\textbf{Example of corpus-in-context prompting} for the NQ dataset, following~\citet{Lee2024LOFT}. The input is categorized by type, with all types being provided as input to the LCLM for retrieval.}
    \label{tab:prompt}
    \vspace{-0.1in}
    \resizebox{1\textwidth}{!}{
    \renewcommand{\arraystretch}{1.5}
        \begin{tabular}{ll}
        \toprule
        \multicolumn{1}{p{.14\textwidth}}{\textbf{Types}} & \makecell{\multicolumn{1}{p{.86\textwidth}}{}} \\
        \midrule
        \midrule
        \multicolumn{1}{p{.14\textwidth}}{Instruction} & 
        \makecell{
            \multicolumn{1}{p{.86\textwidth}}{You will be given a list of documents. You need to read carefully and understand all of them. Then you will be given a query, and your goal is to find all documents from the list that can help answer the query. Print out the ID and TITLE of each document.}\\
            \multicolumn{1}{p{.86\textwidth}}{}\\
            \multicolumn{1}{p{.86\textwidth}}{Your final answer should be a list of IDs, in the following format:}\\
            \multicolumn{1}{p{.86\textwidth}}{Final Answer: [id1, id2, ...]}\\
            \multicolumn{1}{p{.86\textwidth}}{If there is only one ID, it should be in the format:}\\
            \multicolumn{1}{p{.86\textwidth}}{Final Answer: [id1]}\\
            \multicolumn{1}{p{.86\textwidth}}{}\\
            \multicolumn{1}{p{.86\textwidth}}{If there is no perfect answer output the closest one. Do not give an empty final answer.}\\
        }\\
        \noalign{\vskip 0.25ex}\cdashline{1-2}\noalign{\vskip 0.75ex}
        \multicolumn{1}{p{.14\textwidth}}{Corpus \;\;\;\;\;\;\;\;\;\; Formatting} & 
        \makecell{
            \multicolumn{1}{p{.86\textwidth}}{ID: 0 | TITLE: English compound | CONTENT: Major style guides advise consulting a dictionary to determine whether … | END ID: 0}\\
            \multicolumn{1}{p{.86\textwidth}}{ID: 1 | TITLE: The Lord of the Rings: The Return of the King | CONTENT: The music was composed by Howard Shore … | END ID: 1}\\
            \multicolumn{1}{p{.86\textwidth}}{…} \\
            \multicolumn{1}{p{.86\textwidth}}{ID: 881 | TITLE: Dexter (season 3) | CONTENT: While stalking a murderous drug dealer … | END ID: 881} \\
            \multicolumn{1}{p{.86\textwidth}}{ID: 882 | TITLE: Interstellar medium | CONTENT: In the series of investigations … | END ID: 882} 
        } \\
        \noalign{\vskip 0.25ex}\cdashline{1-2}\noalign{\vskip 0.75ex}
        \multicolumn{1}{p{.14\textwidth}}{Few-shot \;\;\;\;\;\;\;\;\;\; Examples} & 
        \makecell{
            \multicolumn{1}{p{.86\textwidth}}{====== Example 1 ======}\\
            \multicolumn{1}{p{.86\textwidth}}{Which document is most relevant to answer the query? Print out the TITLE and ID of the document. Then format the IDs into a list.} \\
            \multicolumn{1}{p{.86\textwidth}}{If there is no perfect answer output the closest one. Do not give an empty final answer.} \\
            \multicolumn{1}{p{.86\textwidth}}{query: where did the dewey decimal system come from} \\
            \multicolumn{1}{p{.86\textwidth}}{The following documents can help answer the query:}\\      
            \multicolumn{1}{p{.86\textwidth}}{TITLE: Dewey Decimal Classification | ID: 199}\\    
            \multicolumn{1}{p{.86\textwidth}}{Final Answer: ['199']}\\   
            \multicolumn{1}{p{.86\textwidth}}{…} \\
        } \\
        \noalign{\vskip 0.25ex}\cdashline{1-2}\noalign{\vskip 0.75ex}
        \multicolumn{1}{p{.14\textwidth}}{Query Formatting} & 
        \makecell{
            \multicolumn{1}{p{.86\textwidth}}{====== Now let's start! ======}\\
            \multicolumn{1}{p{.86\textwidth}}{Which document is most relevant to answer the query? Print out the TITLE and ID of the document. Then format the IDs into a list.}  \\ 
            \multicolumn{1}{p{.86\textwidth}}{If there is no perfect answer output the closest one. Do not give an empty final answer.} \\
            \multicolumn{1}{p{.86\textwidth}}{query: when does monday night raw come on hulu} \\
            \multicolumn{1}{p{.86\textwidth}}{The following documents can help answer the query:}\\         
        } \\
        \bottomrule
        \end{tabular}
    }
\end{table*}

\end{document}